# Electroactive morphing effects on the aerodynamic performance through wobulation around an A320 wing with vibrating trailing edge at high Reynolds number

By

C. Rouaix[1,4], C. Jimenez-Navarro[1], M. Carvalho[2,1,4], C. Raibaudo[5,6], J. Abou-Khalil[2,1], A. Marouf[1,3], Y. Hoarau[3], G. Harran[1], J. Hunt[7], H. Hangan[4], J.F. Rouchon[2], M. Braza[1]

[1]IMFT - Institut de Mécanique des Fluides de Toulouse *UMR 5502 CNRS-INPT-UT3*, Toulouse, FRANCE
[2]LAPLACE - Laboratoire Plasma et Conversion d'Énergie *UMR 5213 CNRS-INPT-UT3*, Toulouse, FRANCE
[3]ICUBE - Laboratoire des sciences de l'ingénieur, de l'informatique et de l'imagerie*, UMR 7357 CNRS-Univ. Strasbourg-ENGEES-INSA*, Strasbourg, FRANCE
[4]Ontario Tech University, Oshawa, CANADA
[5]ONERA - Toulouse, FRANCE
[6]Université d'Orléans, INSA-CVL, PRISME, EA 4229, Orléans, FRANCE
[7]Trinity College, Cambridge, UK

## Abstract

This study aims to investigate the effects of electroactive morphing on a $70cm$ chord A320 wing by means of near trailing edge slight deformation and vibration. Wing morphing is performed by Macro Fiber Composites (MFC) mini-piezoelectric actuators distributed along the span of the "Reduced Scale" (RS) A320 prototype of the H2020 N° 723402 European research project SMS, "Smart Morphing and Sensing for aeronautical configurations", (https://cordis.europa.eu/project/id/723402 and www.smartwing.org/SMS/EU). The configuration studied corresponds to a low-subsonic regime (Mach number 0.063) with a 10° incidence and a Reynolds number of 1 Million. The numerical simulations are carried out with the Navier-Stokes Multi-Block (NSMB) solver, which takes into account the deformation of the rear part of the wing implemented experimentally with the piezoelectric actuators. A detailed physical analysis of the morphing effects on the wake dynamics and on the aerodynamic performance is conducted with a constant amplitude of $0.7mm$ over a wide range of actuation frequencies $[10-600]Hz$. Optimal vibration ranges of $[180-192]Hz$ and $[205-215]Hz$ were found to respectively provide a 1% drag reduction and a 2% lift-to-drag ratio increase compared to the non-morphing (static) configuration. The natural frequencies associated with the shear layer Kelvin-Helmholtz (KH) vortices and the Von-Kármán (VK) vortex shedding were found to play a central role in the modification of the wake dynamics by morphing as well as in the increase of the aerodynamic performance. Actuating at (or close to) the upper shear layer (USL) natural frequency ($\sim 185Hz$) provides an order of 1% drag reduction and 1% lift-to-drag ratio increase, while actuating at (or close to) the lower shear layer (LSL) natural frequency ($\sim 208Hz$) provides an order of 8% lift increase and 2% lift-to-drag increase. Furthermore, the linear variation of the actuation frequency over time, called wobulation, was shown to have significant effects. This approach demonstrated, through an appropriate mapping, the ability to quickly and efficiently detect optimal constant actuation frequency ranges providing aerodynamic performance increase and simultaneously reducing the amplitude of the main instability modes.

# 1) Introduction

The first powered and heavier-than-air airplane, which performed a short flight in 1890 (named "Eole"), designed and built by Clément Ader in the city of Muret near Toulouse, France, was bio-inspired by birds and had deformable wings (Ader 1907). Deformable wings were also part of the Wright Brothers' first airplane in 1903, which opened the epic of aviation. Although airplanes in the last decades have implemented rigid wings and flaps, several recent studies have been devoted to deformable and disruptive wing designs. For example, the European research projects SARISTU (https://cordis.europa.eu/project/id/284562) and Cleansky2 (https://www.clean-aviation.eu/clean-sky-2) have work packages dealing with morphing wings ensuring cambering and including trailing edge devices, (Dimino, et al. 2016) and (Pecora, Amoroso and Magnifico 2016). In the European project AFLoNext, (http://www.aflonext.eu/), a hybrid laminar wing design with a Krüger concept of the leading edge compensated the loss of 5% lift due to wing-pylon interaction. In this project, AFC (Active Flow Control) on the trailing edge of the wing provided a 1-2% fuel savings, while HLFC (Hybrid Laminar Flow Control technology) resulted in 9% fuel savings. Another trailing edge morphing concept, called Adaptive Compliant Trailing Edge, was developed by NASA in cooperation with FlexSys Inc (http://www.flxsys.com/). Numerous studies, including for instance (Lyu and Martins 2015), (Finchman and Friswell 2015) and (Ting, et al. 2018), have demonstrated the efficiency of trailing edge variable-camber control in improving aircraft performance. Such technology has been shown to be effective for a realistic supercritical wing, reducing wave drag and form drag over a wide range of parameters (Niu, et al. 2020). More recently, AIRBUS - UpNext division in Toulouse studied a new wing design inspired by the albatross bird, and implemented a flexible and foldable wing tip (https://www.airbus.com/en/innovation/disruptive-concepts/biomimicry/wings/albatrossone).

Bio-inspired by birds actively adapting their wings to a wide variety of flow conditions (Lentink, et al. 2007), the concept of morphing has become the center of interest for many researchers around the world, (Barbarino, et al. 2011), (Ajaj, Beaverstock and Friswell 2016), (Li, et al. 2018) and (Chu, et al. 2022). Following million years of evolution, birds skillfully manipulate the surrounding turbulence through their sensing system, allowing them to capture pressure fluctuations. As a result, they are able to instantaneously deform their wing camber and vibrate different types of their winglets and feathers along the span to optimize aerodynamic performance in real time while reducing aerodynamic noise. They are able to control flow separation and vortex shedding (Maybury and Rayner 2001) and are capable of extreme maneuverability, including in rapidly changing and gusty environments, while maintaining remarkable aerodynamic efficiency, (Harvey and Inman 2021) and (Harvey, et al. 2022). The most fascinating examples are the large wingspans birds such as the bald eagle's ability to enhance aerodynamic performance by delaying flow separation (Carruthers, Thomas and Taylor 2007) as well as the owl's aptitude to achieve a virtually silent flight (Sarradj, Fritzche and Geyer 2011), (Jaworski and Peake 2013) and (Jaworski et Peake 2020). Morphing is ubiquitous in nature as it is observed not only in birds, but also in bats (Yu and Guan 2015) or in various types of marine animals, such as cephalopods or fishes, which are able to achieve impressive and rapid maneuvers by employing very large body and shape deformations, (Huffard 2006) (Triantafyllou, Weymouth and Miao 2016). Based on the definition given by (McGowan, et al. 2002), morphing in this study refers to the continuous shape adaptation and vibrational behavior of the aerodynamic surface in real time.

Since 2010, the multidisciplinary research team composed of the IMFT and LAPLACE laboratories has been investigating disruptive future wing design through *electroactive* morphing concepts, partly bio-inspired, able to deform and vibrate strategic parts of the wing on multiple time and length scales through suitable electrical actuators. The use of such actuators provides a much lighter and less energy consuming morphing system than mechanical actuators and is in line with the objective of "More Electric Aircraft" - MEA, a priority in the aeronautics industry nowadays and towards Horizon 2050 (Commission européenne 2012). Over the last thirteen years, the IMFT-LAPLACE research team has created the synergistic "Smartwing Morphing Centre" (www.smartwing.org), a platform that brings together six research institutes in Toulouse and that is coordinated by IMFT in close collaboration with AIRBUS ETCT - "Emerging Technologies and Concepts Toulouse", currently called UpNext. The activities of this platform are focused on novel morphing concepts with the goal of significantly improving aerodynamic performance and reducing noise. The aforementioned research platform has received considerable support from the French Foundation "Sciences et Technologies pour l'Aéronautique et l'Espace" - (STAE) (http://www.fondation-stae.net/) - RTRA ("Réseau Thématique de Recherches Avancées") through the federative projects EMMAV - "Electroactive Morphing for Micro-Airvehicles" and DYNAMORPH - "DYNAmic regime electroactive MORPHing".

In the EMMAV project, (Chinaud, Boussaid, et al. 2012) and (Chinaud, Rouchon, et al. 2014) studied large deformations of a plate using Shape Memory Alloys (SMA). The influence of cambering on the turbulence stress tensor and the shear layer instability mode was evaluated for different electric current intensities. In (Scheller, Chinaud, et al. 2015), push-push piezoelectric actuators were used to vibrate the near trailing edge region of a NACA0012 wing for an incidence of $\alpha = 10°$ and a Reynolds number of $Re = 2 \times 10^5$ with a constant actuation frequency in the range of $f_a = [30\text{-}90]Hz$. A significant vortex breakdown of the Kelvin-Helmholtz (KH) eddies in the shear layers past the wing was achieved, resulting in a significant thinning of the near wake and a strong attenuation of the shear layer instability amplitude, known to be a major source of noise. These experiments were performed using Time Resolved Particle Image Velocimetry (TRPIV). The study demonstrated a significant improvement in lift, together with a reduction in drag and noise sources for an optimal actuation of $f_a = 60Hz$. The "hybrid electroactive morphing" concept was introduced in the context of the DYNAMORPH project, (Chinaud, Scheller, et al. 2013). Actuations at different time and length scales were obtained for efficient manipulation. (Scheller 2015) and (Scheller, Jodin, et al. 2016) applied the hybrid morphing on a NACA4412 wing in the Reynolds number range of $Re = [2 \times 10^5 \text{ - } 4 \times 10^5]$. This hybrid morphing was achieved by simultaneously combining SMA and Macro Fiber Composites (MFC) piezoelectric actuators in the near trailing edge region. SMA actuation provided relatively large deformations (of the order of 10% of the actuated chord) over a significant part of the wing at low frequencies (of the order of $1Hz$), while MFC produced relatively small deformations (of the order of 1-2$mm$) at higher frequencies (of the order of $f = [50\text{-}500]Hz$).

In the framework of the SMS project, (Jodin, Motta, et al. 2017) and (Simiriotis, Jodin, et al. 2019) experimentally and numerically investigated the effects of morphing by means of a Higher Frequency

Vibration Trailing Edge (HFVTE) concept applied to the so-called RS A320 wing prototype, with a chord of $c = 70cm$, an incidence of $\alpha = 10°$, and for Reynolds numbers $Re = 5 \times 10^5$ and $Re = 1 \times 10^6$. An extensive parametric study was conducted regarding the trailing edge actuation frequency and amplitude in the range of $f_a = [50\text{-}500]Hz$ and $a_p = [0.5\text{-}4]mm$ respectively. It was shown that amplitudes higher than $3mm$ are detrimental for the aerodynamic performance as they produce significant increase in drag (Simiriotis 2020). The actuation frequency was identified as a key factor in the manipulation of the turbulent wake dynamics, which was significantly impacted when the actuation frequencies were close to or above the natural frequencies of the separated shear layers and of the Von-Kármán mode further downstream. The aforementioned studies also demonstrated the strong feedback effects of this type of morphing on the pressure distribution all over the wing, resulting in considerable aerodynamic performance benefits. The vortex breakdown suppressed the harmful vortices and enhanced those that contribute to increased circulation, similar to the feather movement of large-span hunting birds with a resulting increase in lift (Jodin, Motta, et al. 2017). More recently, (Marouf, Hoarau, et al. 2023) numerically investigated the same configuration through 3D simulations and analysed the constant actuation frequency at $f_a = 300Hz$, a value that was also found to be optimal in the case of the 2D simulations. This study showed that morphing significantly attenuates the three-dimensional character of the coherent structures by virtually suppressing the secondary instability that naturally amplifies around wings beyond a critical incidence (Hoarau, Braza and Ventikos, et al. 2003) and producing a quasi-2D wake. The effects of trailing edge actuation frequency and amplitude have also been numerically investigated in cruise conditions for transonic regimes around the so-called tRS -"transonic regime Reduced Scale" A320 prototype of the SMS project by (Tô, et al. 2019). This study demonstrated that the buffet instability can be controlled by the trailing-edge morphing at selected frequencies capable of producing a "lock-in" effect of the buffet's natural frequency with the actuation frequency. Actuation ranges of $f_a = [300\text{-}350]Hz$ and $f_a = [700\text{-}720]Hz$ were found effective in enhancing this effect, resulting in a significant drag reduction (of the order of 9%) compared to the non-morphing configuration (Braza, Auteri, et al. 2023). A similar lock-in effect was reported by (Zhang, et al. 2021) who performed numerical simulations of a vibrating trailing edge and obtained a drag reduction of 1.66-2.32%.

(Abdessemed, Bouferrouk and Yao 2021) performed a numerical study of the aeroacoustic gains of a harmonically morphing trailing edge at constant frequency, demonstrating a noise reduction of $1.5dB$ for the main instability in the flow and up to $10dB$ for the first superharmonic. By applying near trailing edge slight deformation and vibration for a high-lift wing configuration at different incidences and for $Re = 1 \times 10^6$ and $Re = 2.25 \times 10^6$, (Marouf, Bmegaptche-Tekap, et al. 2021) obtained optimal actuation amplitudes and frequencies. These optimal morphing parameters were found to attenuate the predominant natural shear layer frequency and to delay the formation of Von-Kármán vortices in the wake. Noise reduction of low instabilities was achieved, with a $15dB$ reduction for the predominant frequency peak. (Watkins and Bouferrouk 2022) showed an 11% broadband noise reduction over the entire frequency spectrum investigated, with a 20% average noise reduction at higher frequencies, and a reduced flow separation with the use of the continuous morphed trailing edge compared to the conventional flap. In all of these studies, the trailing edge actuation amplitude and frequency were held constant.

Natural frequencies related to flow dynamics are central to understanding the present wake structure, which is governed by the two separated shear layers. Shear layers have been widely investigated for the excited mixing layer configuration, (Oster and Wygnanski 1982) and (Kourta, et al. 1987). Indeed, when the wake is excited at, or close to, a shear layer natural frequency or its subharmonics, complex non-linear merging phenomena are involved producing significant manipulation of the mixing layer spreading rate, (Ho and Huang 1982) and (Ho and Huerre 1984). (Yamagishi and Tashiro 2002) also found a significant change in the wake vortex structure dynamics for a flat plate with an incidence by introducing a periodic external fluctuation with a frequency equal to the separated shear layer as well as a frequency similar to the vortex shedding. In the same way, morphing can considerably modify the wake dynamics by introducing periodic excitations through slight vibrations of the trailing edge at a constant frequency.

The time modulation (wobulation) of the trailing edge actuation frequency has been experimentally investigated for the first time with respect to morphing in the context of the SMS project, (Jodin, Carvalho, et al. 2023), with the aim of developing a closed-loop feedback controller to increase the aerodynamic performance in real time. A linear sweep of the frequency was applied in open-loop to model the transfer function of the dynamic system used in the feedback controller. In the achieved closed-loop morphing system, the frequency and amplitude were varied to minimize a cost function defined as the variance (*rms*) of the wall pressure signals at selected sensing locations. This method of frequency sweep excitation for identification has been used in various contexts, such as, among others, rotordynamic, (Gu, et al. 2021). The term wobulation has been widely employed in image display systems (Kim, et al. 2016), in radio and radar wave technologies (Gao and Tian 2015), in material properties measurement (Haraoubia, Meury and Le Traon 1990) and (Witoś 2008), in nuclear magnetic resonance spectroscopy (Ferrand, et al. 2015), in signal processing (Koshelev and Andrevej 2016), as well as in the field of acoustic (Zaitseva, et al. 2017). In the context of this work, wobulation corresponds to a linear sweep of the trailing edge actuation frequency, characterized by an initial frequency and a constant frequency increment (or rate of wobulation, in $Hz\ s^{-1}$).

The aims of this study are threefold. The first goal is to analyse the flow dynamics of the non-morphing (static) configuration, that is the reference case, and to evaluate the main natural frequencies at stake. The study is carried out on the $70cm$ chord of the RS prototype at an incidence of $\alpha = 10°$, a Reynolds number of $Re = 1 \times 10^6$ and in the low-subsonic regime (Mach number $Ma = 0.063$). Then, the effects of morphing achieved by near trailing edge slight deformation at a constant actuation frequency are investigated on the wake dynamics, the separation, the wall pressure and the aerodynamic performance. Finally, the last objective of this work is to understand the effects of the linear frequency variation over time (wobulation) on the flow dynamics and how it can be used towards a fast and efficient detection of optimal constant actuation frequency ranges. This last goal is related to ongoing studies of the present research group to optimize the closed-loop controller of the SMS project to allow real time application of optimal actuation frequencies. To the best knowledge of the authors, wobulation is new in the context of morphing design in aeronautics. Therefore, this research represents a first step in the state-of-the-art to investigate the aerodynamic performance increase by this concept.

According to these objectives, section 2 presents the experimental setup with the main physical parameters involved. Section 3 describes in detail the numerical approach and the turbulence modelling used. Section 4 presents the main results. Section 4.1 analyses the unsteady nature of the non-morphing (static) case, the associated instabilities and coherent structures. This part of the study evaluates the natural frequencies and provides a first comparison between experiments and numerical simulations. Section 4.2 examines the morphing effects at constant actuation frequency. The impact on the aerodynamic performance is discussed after performing a large parametric study in the frequency range $f_a = [10\text{-}600]Hz$. Optimal actuations with significant benefits (lift increase, drag reduction) are pointed out. Section 4.3 studies the effects of wobulation. The relation between wobulation and constant actuation is determined based on a time-frequency analysis using wavelets. An appropriate mapping of the fluctuating forces is proposed, which allows the identification of optimal constant actuation frequency ranges. Finally, a parametric study of the initial frequency and frequency increment is carried out to assess the impact of these parameters on the dynamic system. Section 5 concludes the present study and suggests perspectives, especially for the closed-loop implementation.

## 2) Physical parameters and experimental setup

The experiments were performed in the low-subsonic wind tunnel S4 of IMFT with an upstream velocity of $U_\infty = 21.5 m/s$, corresponding to Reynolds and Mach numbers of $Re = 1 \times 10^6$ and $Ma = 0.063$, respectively. The RS A320 morphing prototype has a chord of $c = 70cm$ and a span of $s = 60cm$ (figure 1a). The wing is mounted with an angle of attack of $\alpha = 10°$, which is representative of a typical take-off configuration, in the S4 subsonic wind tunnel (figure 2a) characterized by a 0.1%. inlet turbulence intensity level.

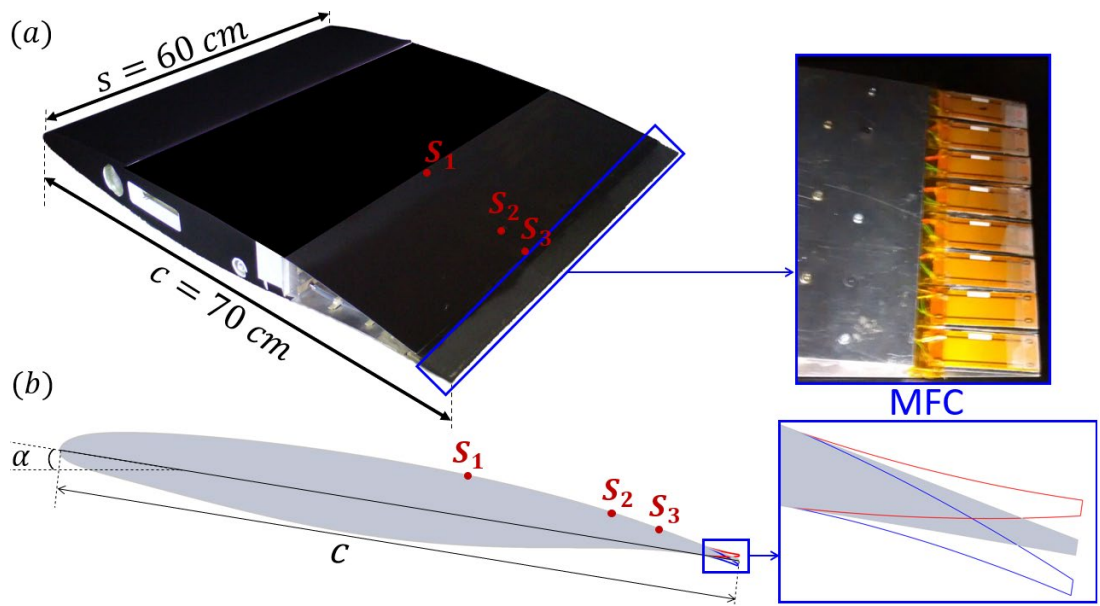

Figure 1 - (a): A320 RS prototype from the SMS project with embedded MFC piezoelectric actuators. (b): Supercritical shape of the wing's mid-section and location of the three highly sensitive pressure transducers. $S_1$, $S_2$ and $S_3$ located on the suction side at $z/c = 0.56$ and $x/c = 0.60$, $x/c = 0.80$ and $x/c = 0.85$ respectively.

The test section has a width of $592mm$ and a height of $712mm$. The lift signals were obtained thanks to an aerodynamic balance designed in a collaboration between LAPLACE and IMFT laboratories. A Yokogawa DL850 oscilloscope was used for lift data acquisition. It provides the zero adjustment function and is able to sample data at $f_s = 5kHz$. Unsteady pressure measurements were performed using three MEGGITT 8507C-1 pressure transducers with a sampling rate of $f_s = 6kHz$. The three sensors, $S_1$, $S_2$ and $S_3$ (figure 1), were implemented at strategic locations on the suction side following numerical simulations using the same configuration as the experiments (Simiriotis, Jodin, et al. 2019). The experimental setup, shown in figure 2c, was controlled by a dSPACE MicroLabBox used to record pressure data as well as to generate the input signal for the piezoelectric actuators and for the triggering of lift data acquisition.

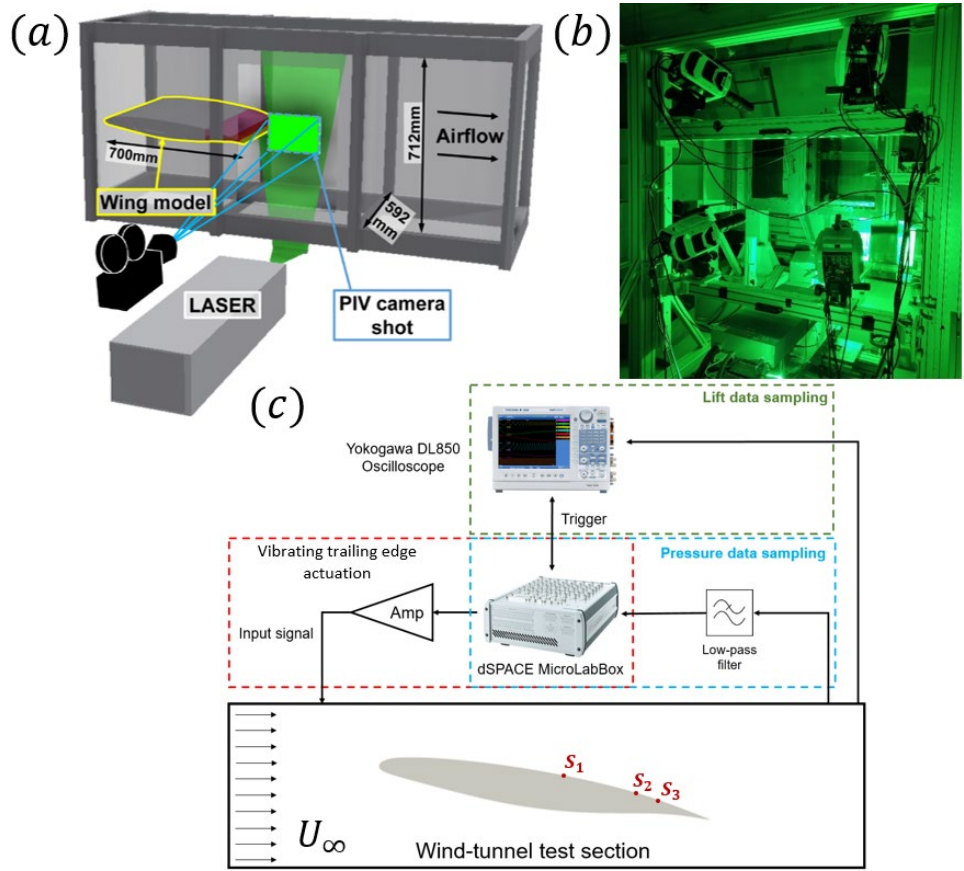

Figure 2 - Experimental setup (a): the subsonic S4 - IMFT wind tunnel test section with the RS A320 wing prototype. (b): the 4D Tomographic PIV with the contribution of "Signaux & Images" team of IMFT. (c): pressure and lift measurements.

The MFC piezoelectric actuators formed patches of length $L_p = 3.5cm$, corresponding to 5% of the chord, and vibration amplitudes up to $a_p = 1mm$ are ensured by a suitable voltage in the range of $[800-1000]V$. These parameters were used according to optimizations from previous studies on this prototype within the SMS project, (Simiriotis, Jodin, et al. 2019) and (Simiriotis 2020).

Regarding the Time-Resolved PIV (TRPIV) campaign, the interrogation window was located downstream of the trailing edge to investigate the shear layers dynamics. A LaVision camera model V2012 with a sampling frequency of $f_s = 10kHz$ was used for image acquisition. The tests were carried out at Reynolds numbers of $Re = 0.7 \times 10^6$ and $Re = 1 \times 10^6$. For particle tracking, correlation between consecutive snapshots was made using $1200 \times 800$ elements, each with a resolution of $16 \times 16$ pixels. LaVision software DaVis and Matlab were used for data processing. During the experiments, the velocity variation was less than 1.5%. For Tomographic PIV, four LaVision V2020 cameras were mounted on a custom-built experimental bench as shown in figure 2b. The sampling frequency was $f_s = 10kHz$, and the Reynolds number was $Re = 5 \times 10^5$. 200,000 snapshots were captured. This Tomographic PIV campaign represents a first 3D velocity field database for morphing wings in the present Reynolds number range.

## 3) Governing equations and numerical method

The numerical simulations were performed using the Navier-Stokes Multi-Block (NSMB) solver, fully parallelized with MPI architectures (Hoarau, Pena, et al. 2016). The governing equations are under the following form:

$$\frac{\partial}{\partial t}(W) + \frac{\partial}{\partial x}(f - f_v) + \frac{\partial}{\partial y}(g - g_v) + \frac{\partial}{\partial z}(h - h_v) = 0 \quad (1)$$

where $t$, $x$, $y$ and $z$ respectively denote time and space cartesian coordinates. The state vector $W$, the convective fluxes $f$, $g$, $h$ and the viscous fluxes $f_v$, $g_v$, $h_v$ are respectively defined as following:

$$W = \begin{Bmatrix} \rho \\ \rho u \\ \rho v \\ \rho w \\ \rho E \end{Bmatrix}, f = \begin{Bmatrix} \rho u \\ \rho u^2 + p \\ \rho uv \\ \rho uw \\ u(\rho E + p) \end{Bmatrix}, g = \begin{Bmatrix} \rho v \\ \rho vu \\ \rho v^2 + p \\ \rho vw \\ v(\rho E + p) \end{Bmatrix}, h = \begin{Bmatrix} \rho w \\ \rho wu \\ \rho wv \\ \rho w^2 + p \\ w(\rho E + p) \end{Bmatrix}$$

$$f_v = \begin{Bmatrix} 0 \\ \tau_{xx} \\ \tau_{xy} \\ \tau_{xz} \\ (\tau U)_x - q_x \end{Bmatrix}, g_v = \begin{Bmatrix} 0 \\ \tau_{yx} \\ \tau_{yy} \\ \tau_{yz} \\ (\tau U)_y - q_y \end{Bmatrix}, h_v = \begin{Bmatrix} 0 \\ \tau_{zx} \\ \tau_{zy} \\ \tau_{zz} \\ (\tau U)_z - q_z \end{Bmatrix}$$

where $\rho$ is the density, $u$, $v$ and $w$ correspond to the cartesian velocity components, $p$ is the pressure, $E$ is the total energy, $q_*$ is the heat flux due to conduction calculated according to Fourier's Law and $\tau$ is the shear tensor. The viscous dissipation in the energy equation is calculated from: $(\tau U)_* = \tau_{*x}u + \tau_{*y}v + \tau_{*z}w$.

*Turbulence modelling*

An adapted turbulence modelling approach sensitized to accurately simulate coherent structures development, the Organised Eddy Simulation (OES), (Braza, Perrin and Hoarau 2006), (Bourguet, et al. 2008) and (Szubert, et al. 2015), was employed. This turbulence modelling, based on the phase (ensemble) averaging of the Navier-Stokes equations, consists of a resolved part containing all the organised (or coherent) motions in time and space and a modeled part containing the chaotic (or random) turbulence effects, thus reproducing the dual character of turbulence: organised and chaotic, (Brown and Roshko 1974). This approach is based on the splitting of the turbulence spectrum into one part formed by the organised modes (predominant peaks or frequency bumps) and a second part (continuous spectrum) corresponding to the chaotic processes. This turbulence modelling therefore distinguishes the structures to be resolved from those to be modeled according to their physical nature, organised and chaotic, rather than by their size, allowing reliable simulations in the high Reynolds number range with reasonable grid sizes. The OES approach hence lies in between the Large Eddy Simulation (LES) and the statistical approaches. It was widely employed in the development and improvement of hybrid turbulence modelling since the European project DESIDER, "Detached Eddy Simulation for Industrial Aerodynamics", (Haase, Braza and Revell 2009). It is particularly suitable in the present context of inhomogeneous turbulent flows in non-equilibrium, as it allows to capture the near wall turbulence stress anisotropy, with emphasis on the detached flow regions produced by the relatively high angle of attack and high Reynolds number at stake. The OES approach is non-inherently 3D and can be applied in the scope of 2D simulations, since the main dynamical properties of the coherent structures have a predominantly 2D origin, (Persillon and Braza 1998). As a result, an extensive 2D parametric study was conducted with respect to the trailing edge frequency for constant actuation and with respect to both the initial frequency and the frequency increment for wobulation. This allowed detection of optimal parameter ranges for the vibration to be used in further selective 3D simulations focusing on a narrower set of parameters and providing a deeper physical insight.

*Deformation and vibration of the trailing edge*

The vibration and slight deformation of the near trailing edge region in numerical simulations exactly reflect the actuation obtained with MFC patches in experiments. The motion and deformation of the near trailing edge region produce a mesh deformation, which is taken into account in the NSMB code by the Arbitrary Lagrangian Eulerian (ALE) approach, (Donea, Giuliani and Halleux 1982), where the convection of the mesh nodes is integrated into the transport equations as following:

$$\frac{\partial}{\partial t}\iiint_{\Omega(t)} \vec{W}(x,\tau)d\Omega + \oiint_{\partial\Omega(t)} \vec{H}\vec{n}dS = \vec{0} \quad (2)$$

$$\text{where } \vec{H} = \begin{Bmatrix} \rho(\vec{V}-\dot{\vec{x}})\vec{n} \\ \rho\vec{V}(\vec{V}-\dot{\vec{x}})\vec{n} + p\vec{n} - \bar{\bar{\tau}}\vec{n} \\ \rho E(\vec{V}-\dot{\vec{x}})\vec{n} + p\vec{V}\vec{n} - (\bar{\bar{\tau}}.\vec{V})\vec{n} + \vec{q}\vec{n} \end{Bmatrix}$$

Remeshing methods using Volume Spline Interpolation (VSI) and Trans-Finite Interpolation (TFI) (Guillaume, et al. 2011), were employed in the NSMB code to compute the new deformed grid. The deformed shape of the patches follows a second-order polynomial expression as shown in figure 3:

$$y(x,t) = \left(\frac{2h_p}{3L_p^2}(x-x_p)^2 + \frac{h_p}{3L_p}(x-x_p)\right)\sin(2\pi ft) \quad (3)$$

where $t$ is the time, $y$ is the vertical displacement of the MFC patches, $x$ is the chordwise direction, $h_p = 0.35mm$ is the semi-amplitude at the trailing edge location $x_{TE}$, $L_p = 3.5cm$ (corresponding to 5%c) is the patch length on which the displacement is applied in both experiments and numerical simulations and $x_p$ is the origin of the displacement defined by $x_p = c - L_p$. The trailing edge actuation amplitude $a_p = 2h_p = 0.7mm$ is located at the ending tip of the trailing edge, as presented in figure 3a. $f$ is the trailing edge actuation frequency and can be either constant (or monochromatic) $f = f_a$ or modulated in time by means of wobulation $f = f_w$. In the latter case, the frequency varies linearly as a function of time. All the actuation frequencies investigated are included in the range $f_a = [10\text{-}600]Hz$ in both experimental and numerical approaches.

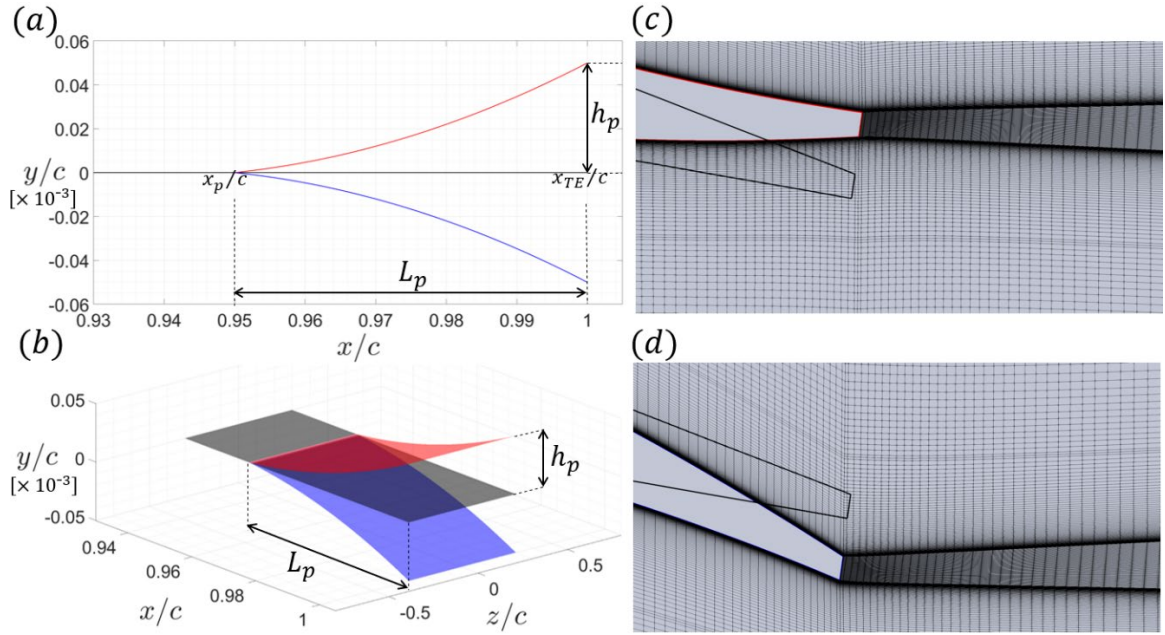

Figure 3 - Vertical displacement $y(x,t)$ in numerical simulations reflecting the MFC piezoelectric vibrations in experiments. (a): 2D and (b): 3D view. Mesh deformation during the upwards deflection (in red) in (c) and the downwards deflection (in blue) in (d). For interpretation of the references to color in this figure legend, the reader is referred to the web version of this article.

*Numerical parameters*

For the temporal discretization, an implicit dual time-stepping with a second order implicit backward difference scheme was employed. After a thorough numerical study (Simiriotis, Jodin, et al. 2019), the outer time step was set to $\Delta t = 1 \times 10^{-5} s$. Advection and diffusion terms were spatially discretized thanks to a fourth order standard central skew-symmetric scheme with artificial second and fourth order dissipation terms (Jameson 1993) and second order central scheme respectively. (Marouf 2020) investigated two different preconditioning methods in order to increase accuracy for the low-subsonic flow. The artificial compressibility, using LU-SGS method (Chorin 1967), was chosen after providing better agreement of the velocity deficit with the experimental results The turbulence intensity corresponds to 1% of the freestream velocity.

*Computational domain and grid parameters*

The computational domain was designed with a rectangular shape based on the height of the S4 wind tunnel, as shown in figure 4a. The 2D structured multi-block mesh using a total of 64 blocks and approximately 300,000 finite volume cells with a quadrilateral shape was selected after a mesh convergence study, which showed good agreement with the experimental results and represented a good compromise between accuracy and reasonable CPU time (Simiriotis, Jodin, et al. 2019). A zoomed view of the mesh is displayed in figure 4b. The height of the first cell layer was chosen accordingly to obtain a $y^+$ of the order of 0.1, to accurately capture the turbulence damping towards the wall as well as the boundary layer dynamics and thickness.

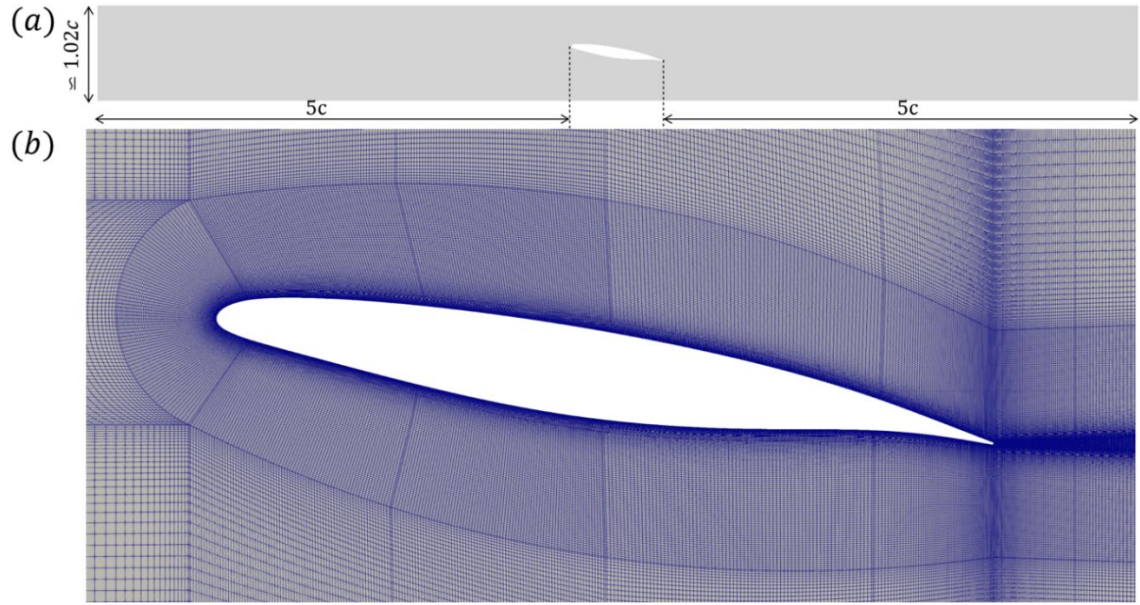

Figure 4 - (a): 2D view of the computational domain for numerical simulations. The dimensions are based on the experiments carried out in the S4 wind tunnel. (b): Zoomed view of the mesh around the A320 airfoil.

*Initial and boundary conditions*

All the cases investigated were initialized from a steady state solution with an upstream velocity of $U_\infty = 21.5 m/s$ and with standard atmospheric conditions of $P_\infty = 101325 Pa$ and $T_\infty = 293.15 K$, corresponding to air pressure and temperature respectively. The effects of the boundary conditions for the upper and lower wind tunnel walls and of their respective boundary layers were investigated through slip and no-slip conditions (Marouf 2020). No-slip conditions were found to provide very small differences compared to the slip conditions. Slipping walls provided a better agreement with the experimental results and were finally chosen for the computations. No-slip conditions were used over the airfoil and a Dirichlet boundary condition on the velocity was imposed at the inlet. Non-reflective conditions were imposed on the outlet boundary (Poinsot and Lele 1992), based on a more general non-reflective formulation (Jin and Braza 1993), inspired by electromagnetic field equations and suitable for low-subsonic or incompressible flow regimes.

## 4) Results

This section first presents the static (non-morphing) case and introduces the natural frequencies in the flow. Then, the effects of morphing are investigated through a large parametric study with respect to constant actuation frequency. The modification of the wake dynamics and the aerodynamic performance of the morphing cases are compared to the static case. Finally, the wobulation of the actuation frequency is examined in terms of its initial frequency and frequency increment, and the relation between constant actuation frequency $f_a$ and wobulation frequency $f_w$ is determined.

## 4.1) Static (non-morphing) case
*Flow dynamics*

Figures 5a and 5b respectively show streaklines visualisation of an instantaneous 2D snapshot of the far wake region and a zoom of the near trailing edge region flow field obtained by numerical simulations for the static reference case ($\alpha = 10°$ and $Re = 1 \times 10^6$). Figure 6 shows the flow dynamics obtained by experiments for the same parameters. In both approaches, the flow comes from the left and goes to the right. The color of each particle represents its initial position and was chosen to clearly distinguish the evolution of the particles coming from the suction side from those coming from the pressure side.

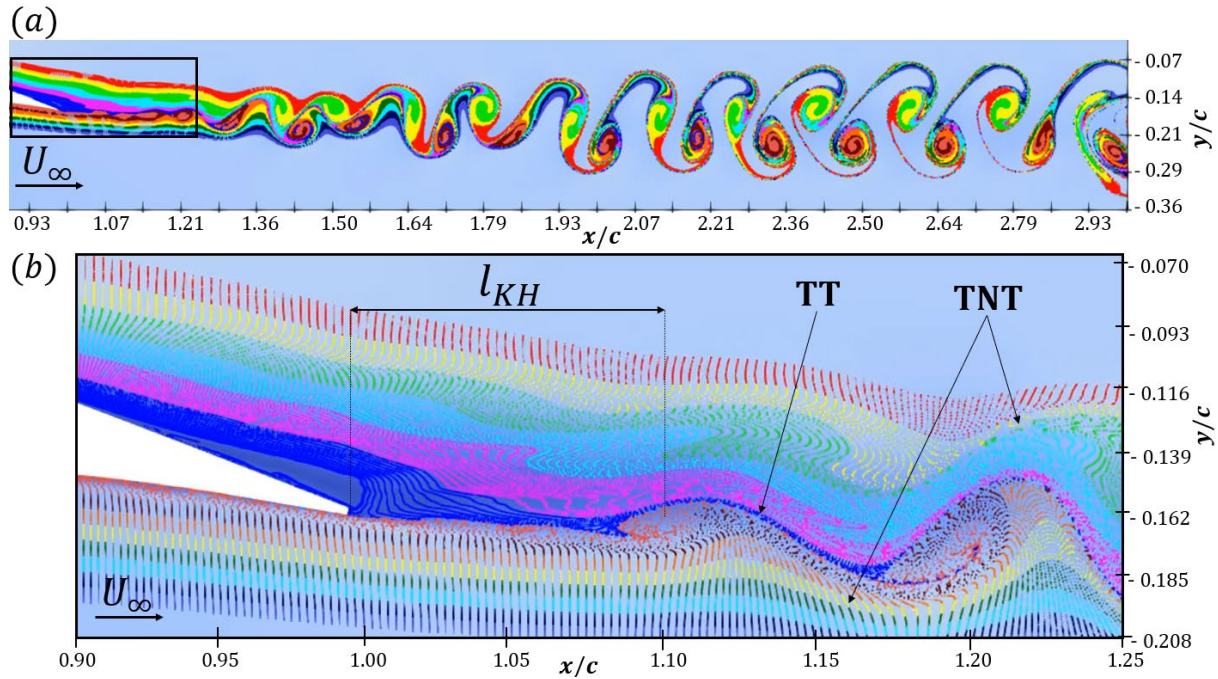

Figure 5 - (a): Streaklines visualisation of the wake for the static case (numerical simulations), (b): zoom of the near trailing edge region. The incidence and Reynolds number are respectively $\alpha = 10°$ and $Re = 1 \times 10^6$. TT and TNT stands for Turbulent/Turbulent interface and Turbulent/Non-Turbulent interface respectively (Hunt, et al. 2016). $l_{KH}$ is the Kelvin-Helmholtz formation length. For interpretation of the references to color in this figure legend, the reader is referred to the web version of this article.

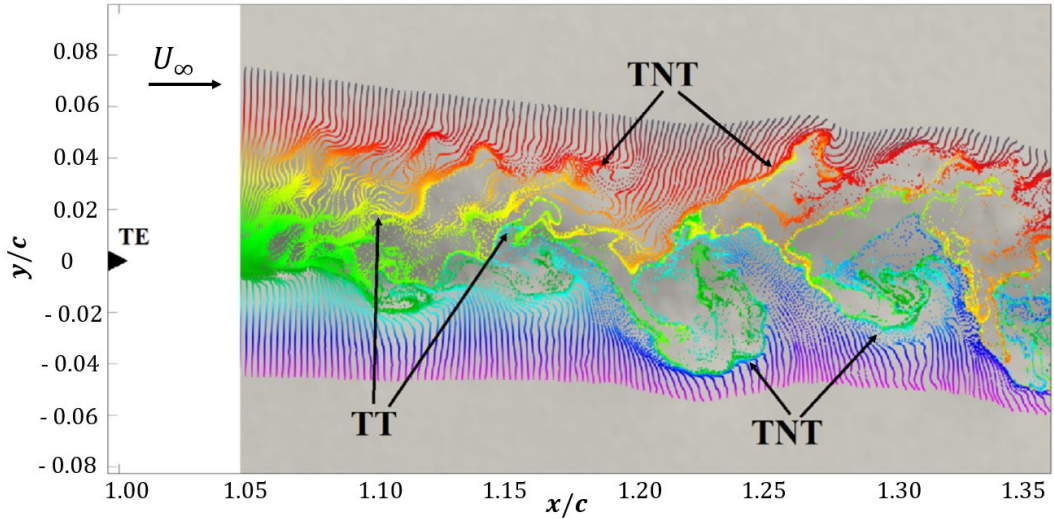

Figure 6 - Streaklines visualisation of the near trailing edge region for the static case by TRPIV (experiments from (Jodin, Motta, et al. 2017)). The incidence and Reynolds number are respectively $\alpha = 10°$ and $Re = 1 \times 10^6$. TT and TNT stands for Turbulent/Turbulent interface and Turbulent/Non-Turbulent interface respectively. For interpretation of the references to color in this figure legend, the reader is referred to the web version of this article.

Figure 7 shows the 3D flow dynamics for numerical simulations of the static case with the same physical parameters as the 2D configuration. The visualisation of the iso-contours of the $Q$ criterion in figure 7a allows to highlight the main wake structures and the formation of secondary instability along the span, (Persillon and Braza 1998), (Braza, Faghani and Persillon 2001) and (Hoarau, Braza and Ventikos, et al. 2003). The configuration of a non-symmetric airfoil with a relatively high incidence and Reynolds number is characterized by a recirculation region on the suction side as shown in figures 5b and 7b for 2D and 3D numerical simulations respectively.

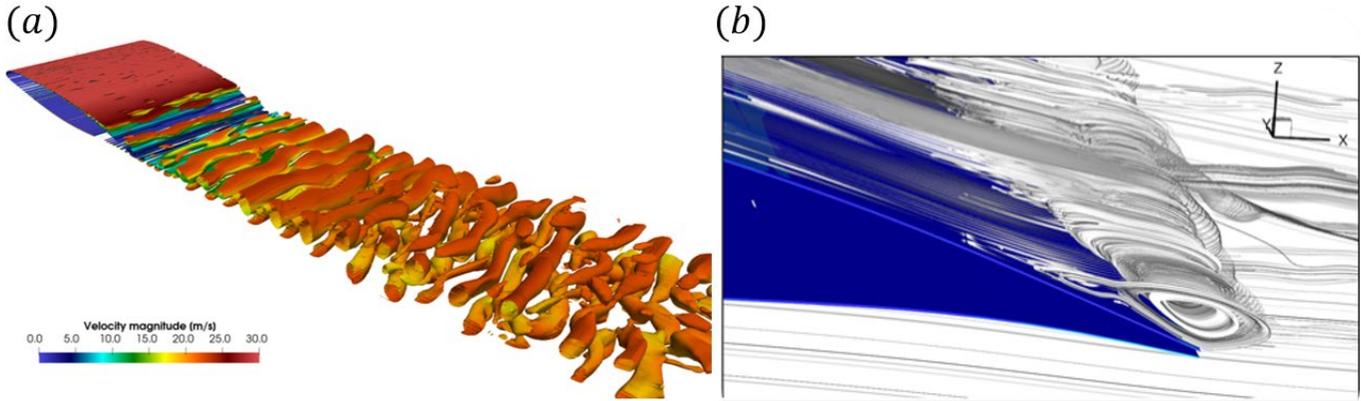

Figure 7 - 3D simulations for the static case: (a) Iso-contours of the $Q$ criterion ($Q = 1000 s^{-2}$) colored by the velocity field magnitude for the static case and (b) Streamlines illustrating the recirculation region (Marouf, Hoarau, et al. 2023). For interpretation of the references to color in this figure legend, the reader is referred to the web version of this article.

As shown in figure 5a, the turbulent wake is formed by two interfacial shear layers (upper and lower) interacting non-linearly with each other in the near trailing edge region (for $x/c = 1.30$) and forming a clear Von-Kármán alternating eddies pattern farther downstream (for $x/c \geq 1.90$). Kelvin Helmholtz (KH) vortices appear in the two separated shear layers after a formation length $l_{KH}$, indicated in figure 5b. Due to the relatively high angle of attack and the supercritical shape of the wing, the effective shear of the lower shear layer (LSL) formed on the pressure side of the airfoil is significantly higher than that of the upper shear layer (USL) formed on the suction side. As a result, the LSL vortices (orange particles) are reinforced and stronger compared to USL eddies (green particles), as illustrated in figures 5a and 5b. By tracking these lower and upper KH vortices through a series of streakline snapshots (similar to the one shown in figure 5a), their respective natural frequencies were evaluated as $f_{LSL} \sim 210 Hz$ and $f_{USL} \sim 185 Hz$ for the lower and upper shear layers respectively. Using the same tracking methodology, an estimation of the Von-Kármán vortex natural shedding frequency was found approximately of the order of $f_{VK} \sim 135 Hz$. These values were accurately determined by means of spectral analysis of the vertical component of the velocity for monitor points located at strategic positions in the wake, as detailed below.

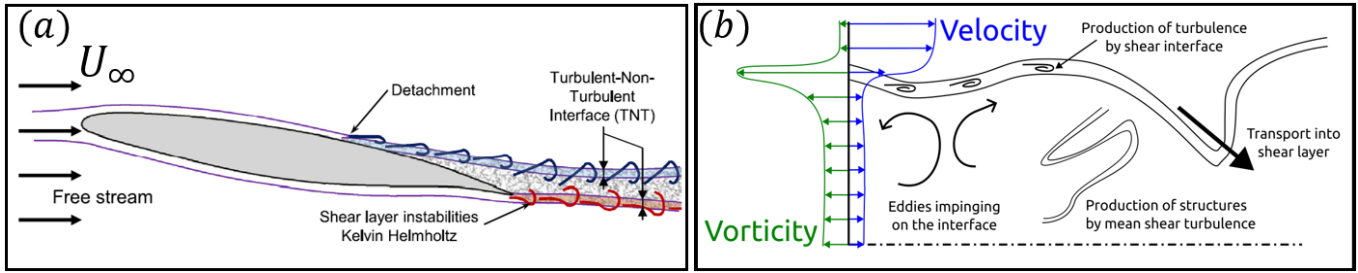

Figure 8 - (a) Sketch of the wake dynamics showing the separated shear layers constituted of KH vortices. (b) Schematic representation of the eddy-blocking effect along the TNT interface (adapted from (Jodin, Motta, et al. 2017)).

The shear layer regions delimit the Turbulent/Non-Turbulent (TNT) interfaces as sketched in figure 8, (Bisset, Hunt and Rogers 2002), (Da Silva, et al. 2014) and (Szubert, et al. 2015). Due to the interaction between the two TNT interfaces, an additional thin inner region, called Turbulent/Turbulent (TT) interface is formed. This region results from the interaction of the upper and lower shear layers with the chaotic turbulence. As discussed in section 4.2, when actuating at optimal frequency and amplitude, morphing is able to efficiently manipulate these interfaces to enhance beneficial vortices while causing the breakdown of harmful ones with respect to the aerodynamic efficiency (lift increase and drag reduction).

*Natural frequencies related to the flow dynamics*

Figure 9 presents the spectral analysis for selected monitoring points that were chosen to determine the natural frequencies previously assessed based on streaklines visualisation (see for instance figure 5). The Power Spectral Density (PSD) of the vertical velocity component fluctuations $v'$ is plotted in figures 9a, 9b and 9c for monitor points located respectively in the recirculation region, along a vertical line in the near wake region, and along an horizontal line in the far wake region. In addition, figure 9d shows the PSD of the lift coefficient fluctuations $C_L'$ providing a global analysis of the flow dynamics. PSDs are computed over signals with a statistical length of about 14 seconds duration using the Welch's overlapped segment averaging estimator (Welch 1967) with a sliding windowing FFT to reduce the variance of the periodogram. Multiple segment sizes were applied from the higher to the lower frequency ranges respectively in order to clearly capture the rich content across the entire energy spectrum. Each segment was filtered by a Hamming window with a 50% overlap between them.

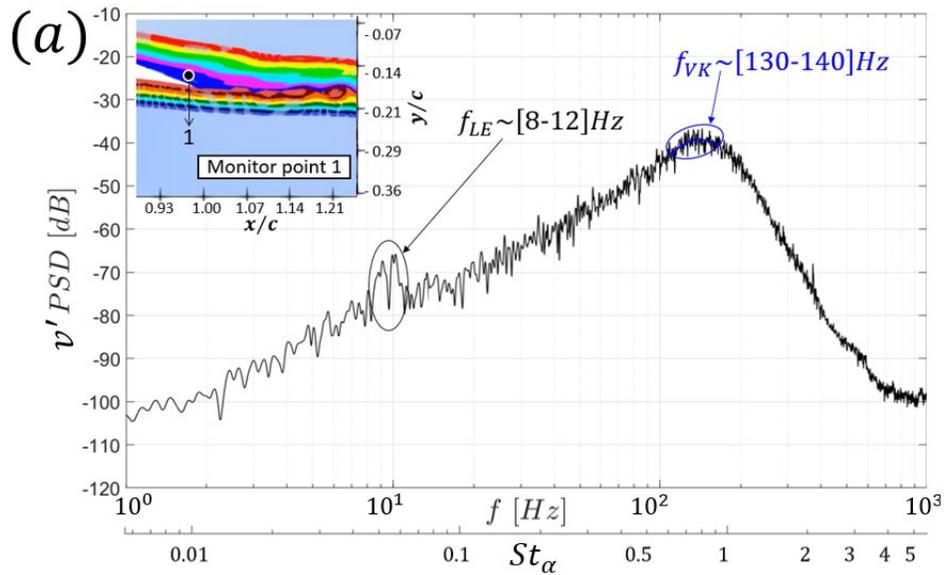

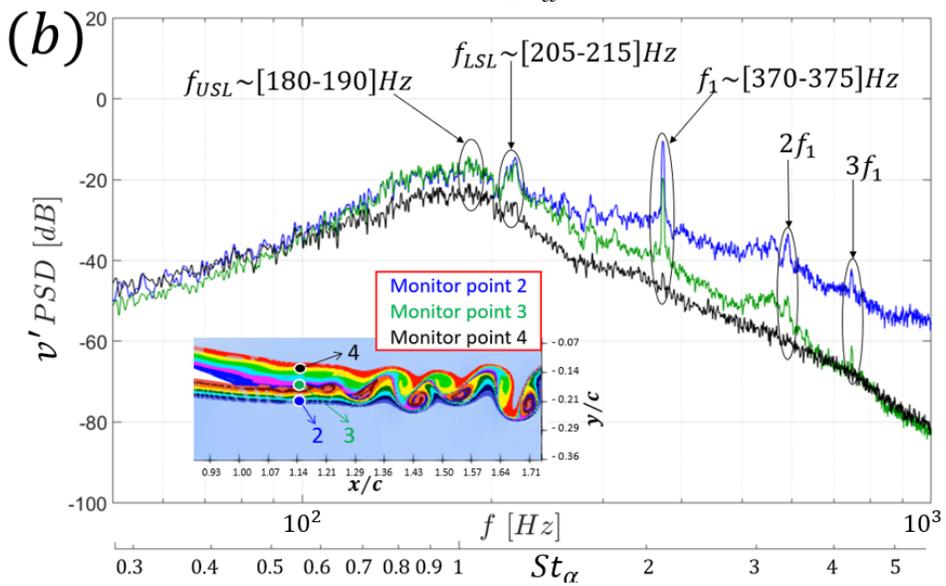

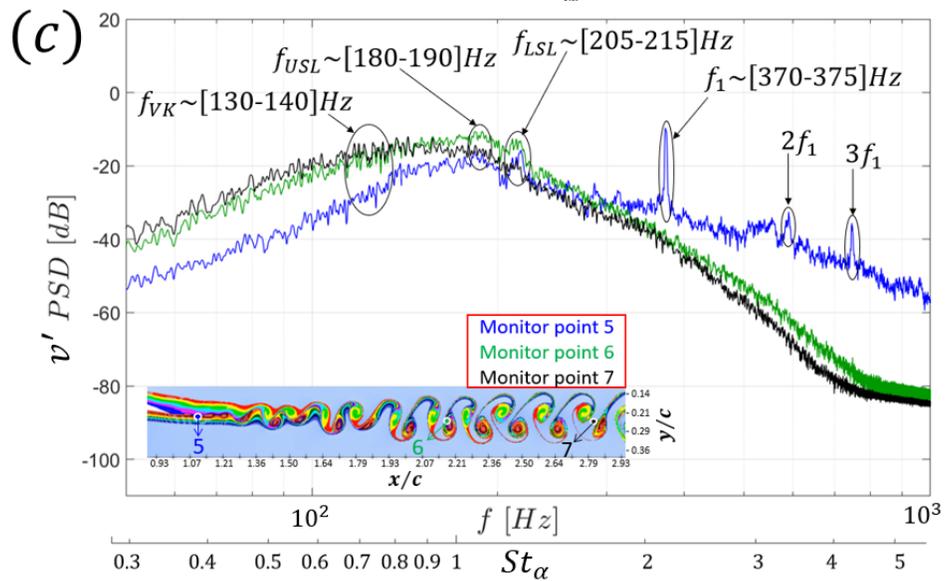

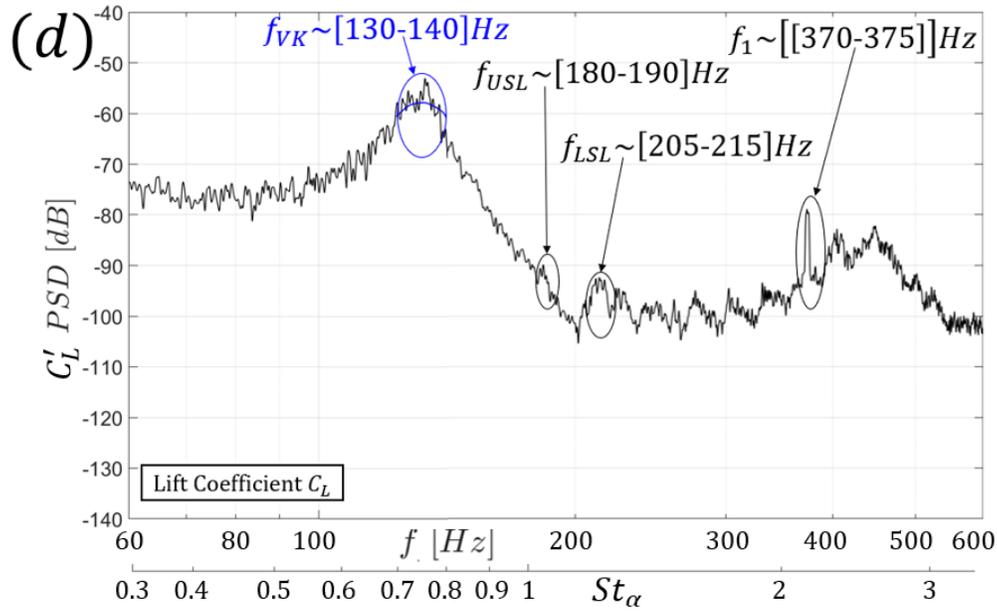

Figure 9 - PSDs of the vertical velocity component fluctuations $v'$ for monitor points located (a): in the recirculation region, (b): along a vertical line, and (c): along an horizontal line. (d): PSD of the lift coefficient fluctuations $C'_L$. $f_{LE}$, $f_{VK}$, $f_{USL}$ and $f_{LSL}$ respectively correspond to leading edge small separation bubble as referred in (Hoarau 2002), Von-Kármán, upper shear layer and lower shear layer natural frequencies. $f_1$ corresponds to an interaction mode which is an n-integer combination of shear layer instabilities. $St_\alpha$ corresponds to the Strouhal number based on the projected area that is $c\sin(\alpha)$. For interpretation of the references to color in this figure legend, the reader is referred to the web version of this article.

|  | $f$ [Hz] | $St = \dfrac{fc}{U_\infty}$ | $St_\alpha = \dfrac{fc\sin(\alpha)}{U_\infty}$ | $St_d = \dfrac{fd_{inst}}{U_\infty}$ |
|---|---|---|---|---|
| Leading Edge small separation bubble: $f_{LE}$ | 8-12 | 0.26-0.39 | 0.045-0.068 | - |
| Vortex Shedding (Von-Kármán): $f_{VK}$ | 130-140 | 4.23-4.56 | 0.74-0.79 | 0.30-0.33 |
| Upper Shear Layer (Kelvin-Helmholtz): $f_{USL}$ | 180-190 | 5.86-6.19 | 1.02-1.07 | 0.17-0.18 |
| Lower Shear Layer (Kelvin-Helmholtz): $f_{LSL}$ | 205-215 | 6.67-7.00 | 1.16-1.22 | 0.19-0.20 |
| $f_1 = \dfrac{n_1 f_{LSL} \pm n_2 f_{USL}}{n_3}$ | 370-375 | 12.05-12.21 | 2.09-2.12 | - |

Table 1 - Natural frequencies for the static case: $f_{LE}$, $f_{VK}$, $f_{USL}$ and $f_{LSL}$ respectively correspond to leading edge small separation bubble, Von-Kármán, upper shear layer and lower shear layer natural frequencies. $f_1$ corresponds to a non-linear interaction mode which is combination of the shear layers instabilities ($n_1$, $n_2$ and $n_3$ are n-integer).

Table 1 summarizes the main natural frequencies in the flow together with their respective Strouhal numbers (or reduced frequencies) based on different characteristic lengths. The nominal Strouhal $St$ number is based on the chord of the airfoil $c$, while $St_\alpha$ and $St_d$ are the Strouhal number based on the projected area $c\sin(\alpha)$ and the dynamic Strouhal number based on the instability mean vortex size $d_{inst}$ in the near trailing edge region for the shear layers and in the far wake for the Von-Kármán alternated eddies respectively.

Due to the relatively high Reynolds number, turbulence produces non-linear interactions among a multitude of chaotic modes and the organised (or coherent) ones (e.g. Von-Kármán and the two shear layers), as well as their sub and superharmonics. Therefore, the overall effect of the organised modes does not appear as a sharp peak in the turbulence spectrum, but rather as a predominant frequency bump spread over a range of frequencies. This bump is due to the relatively high Reynolds number involved, which induces a smearing of the alternating vortices in terms of phase and amplitude irregularities. A clear illustration of this phenomenon is observed in figure 9a where the PSD highlights a prominent spread bump (not a sharp peak) over a frequency range $f = [130\text{-}140] Hz$ with a maximal spectral amplitude close to the Von-Kármán natural frequency $f_{VK}$ previously assessed in section 4.1. The recirculation region is known to oscillate at the Von-Kármán frequency. Indeed, because of the elliptic nature of the flow associated with the low-subsonic regime, perturbations from far downstream in the wake are more prone to propagate upstream and thus can produce beneficial *feedback* effects (Bouhadji and Braza 2003). As a result, the Von-Kármán instability, which is expected to play a significant role in the far wake region, also has a significant influence on the recirculation region and on the overall upstream pressure distribution. This oscillation of the recirculation region drastically affects the lift force, which also displays a prominent bump over the region containing the Von-Kármán natural frequency $f_{VK}$, as shown in figure 9d presenting the PSD of the lift coefficient. As highlighted in table 1, the dynamic Strouhal number based on the mean alternating vortex size ($d_{inst} \sim 5cm$) is $St_d \sim 0.3$. This value is in good agreement with the Strouhal number reported in measurements by (Ribner and Etkin 1958) and reported by (Roshko 1961) for a circular cylinder in the Reynolds number range of $Re = 1.70 \times 10^5$. Using a qualitative analogy with the circular cylinder wake, this range corresponds to an effective Reynolds number of $Re_{eff} = Re\ sin(\alpha) \sim 1.73 \times 10^5$. In agreement with streaklines visualisation from figure 5a showing the clear formation of Von-Kármán alternated eddies in the far wake region, figure 9c shows the enhancement of the spectral density amplitude in the low frequency region, especially around the Von-Kármán frequency bump, for point 7 located far downstream in comparison with point 5 located in the near trailing edge region. Figure 9b highlights the relative importance of each shear layer along a vertical line located in the near trailing edge region: while two distinctive bumps of similar spectral amplitude corresponding to the upper shear layer $f_{USL}$ and the lower shear layer $f_{LSL}$ are visible at the bottom for point 2, only one prominent spread bump of maximal spectral amplitude close to the upper shear layer $f_{USL}$ remains visible at the top for point 4. This shows the dominance of the USL over the LSL as one moves up along a vertical line, as expected. The spectral amplitude of the two shear layer modes is expected to decrease further downstream where the Von-Kármán spectral amplitude increases (figure 9c).

Figures 9b, 9c and 9d also show a predominant sharp peak $f_1$, together with its first and second harmonics, which is a non-linear interaction between several frequencies, producing a new frequency as an n-integer combination of the two initial shear layer frequencies as shown in table 1. The interaction between two (or more) predominant incommensurate frequencies contributes to the filling of the energy spectrum by a multitude of distinct frequencies due to coherent structures and regions of interspersed continuous frequency ranges due to chaotic turbulence. Despite the strong non-linear character of this turbulence-induced chaos, it is known that the newly generated frequencies resulting from the interaction, such as $f_1$ in the present case, can be expressed as a linear expression of the initial predominant frequencies (Newhouse, Ruelle and Takens 1978). Figure 9a shows low frequency modes, assumed to be associated with the oscillation of a small leading edge separation bubble, (Hoarau, Braza and Rodes, et al. 2002), (Hoarau 2002) , (Zaman, McKinzie and Rumsey 1989) and (Yarusevych, Sullivan and Kawall 2009).

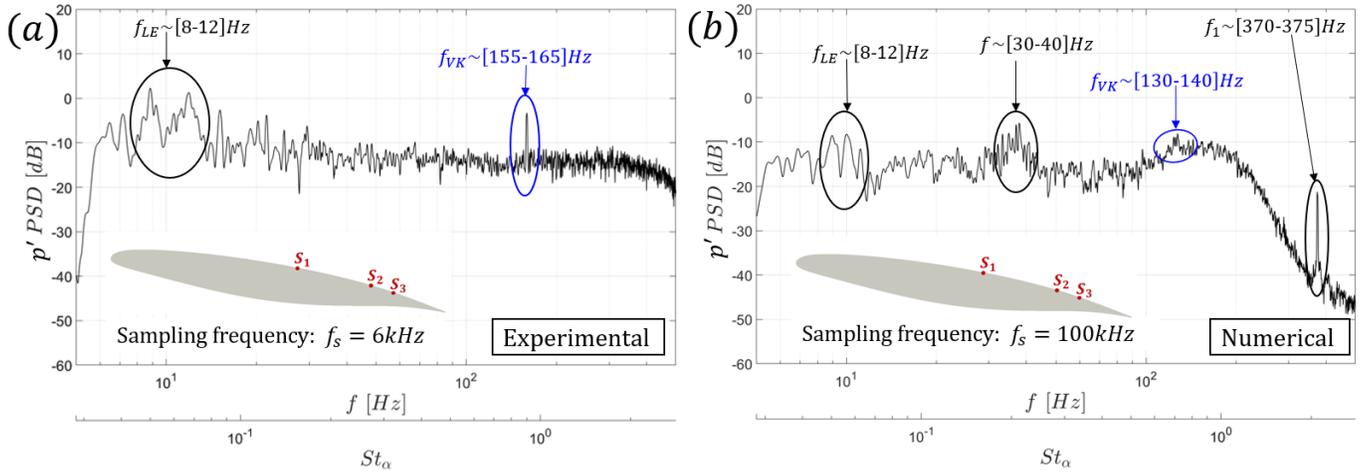

Figure 10 - PSD of pressure fluctuations signals $p'$ on sensor $S_2$ located at $x/c = 0.80$ for the static reference case (14$s$ duration signals, $Re = 1 \times 10^6$) - (a) Experiments and (b) Numerical simulations.

Figures 10a and 10b show PSDs of pressure fluctuation signals for experiments and numerical simulations respectively. The order of magnitude of the spectral amplitude is similar between experiments and numerical simulations. The previously mentioned low frequency bumps $f_{LE}$ are clearly appearing in both experiments and numerical simulations.

## 4.2) Morphing: constant actuation frequency

Figure 11 shows a comparison of a Time-Resolved Tomographic PIV instantaneous snapshot between static and morphing configurations. A significant thinning of the near wake is obtained, as well as a 15% reduction in the recirculation length, which is associated with the drag reduction discussed in the next section.

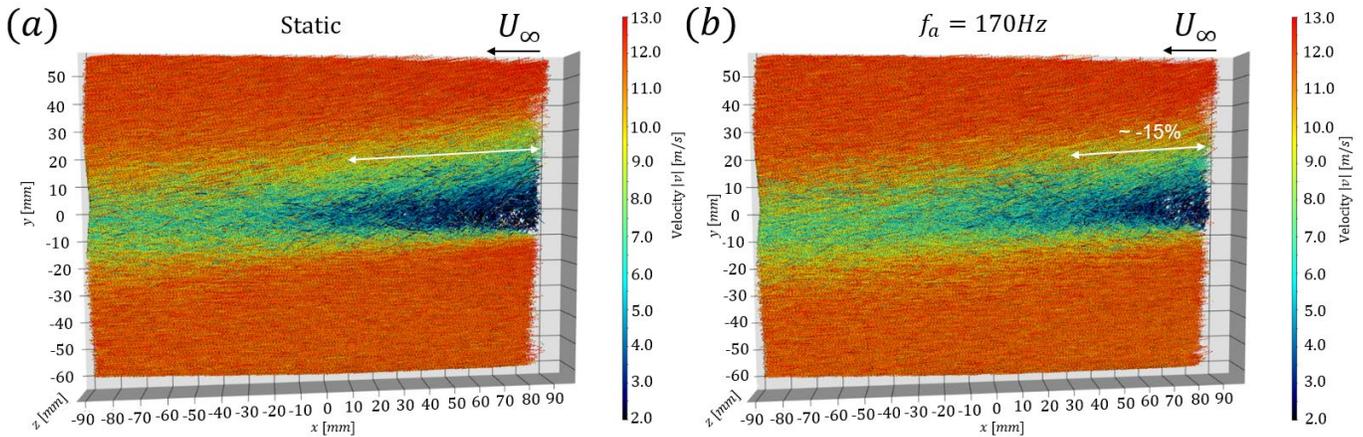

Figure 11 - Instantaneous velocity field obtained through Tomographic PIV (depth of $z = 15mm$) for $Re = 5 \times 10^5$, with the contribution of "Signaux & Images" team of IMFT. (a) static case. (b) constant actuation frequency at $f_a = 170Hz$. The flow comes from the right to the left (Carvalho 2021). For interpretation of the references to color in this figure legend, the reader is referred to the web version of this article.

Figures 12a and 12b show the pressure fluctuation signals for the static case overlapped with one constant vibration frequency case $f_a = 185Hz$ for experiments and numerical simulations respectively. For this actuation case, the order of magnitude of the lift coefficient fluctuations is similar in both experiments and numerical simulations.

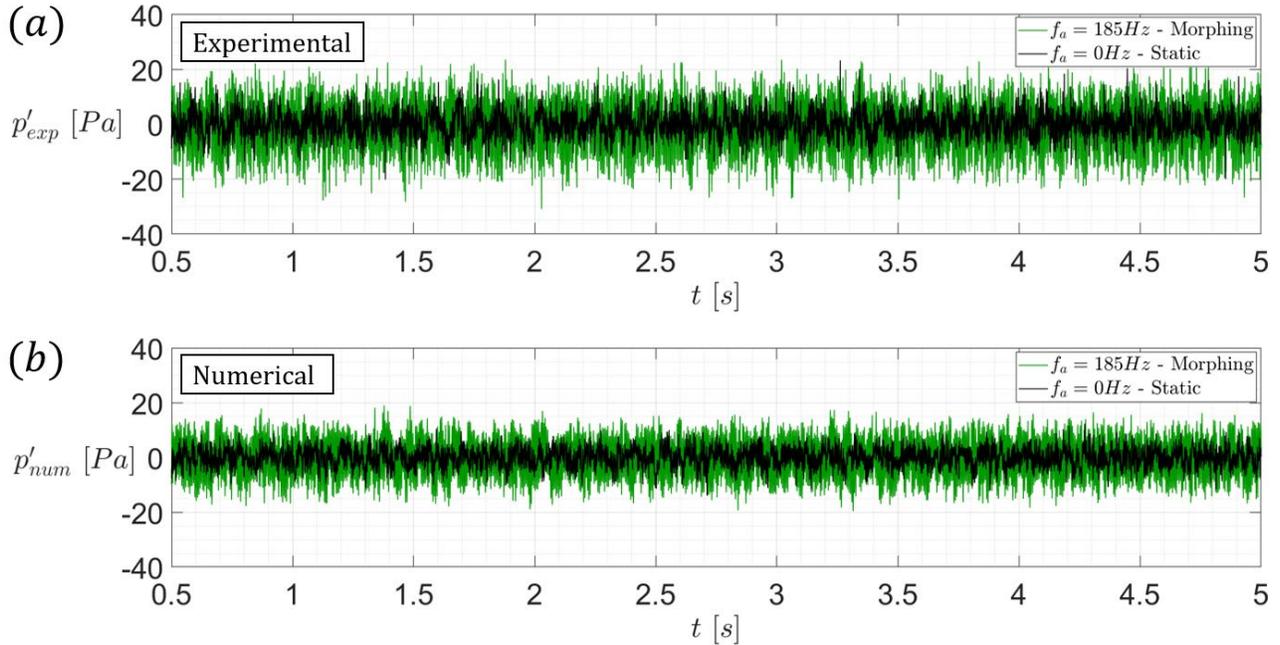

Figure 12 - Pressure fluctuations signals $p'$ for the static reference case and the morphing case (constant actuation frequency at $f_a = 185Hz$) on pressure sensor $S_2$ and for $Re = 1 \times 10^6$ - (a) Experiments and (b) Numerical simulations. For interpretation of the references to color in this figure legend, the reader is referred to the web version of this article.

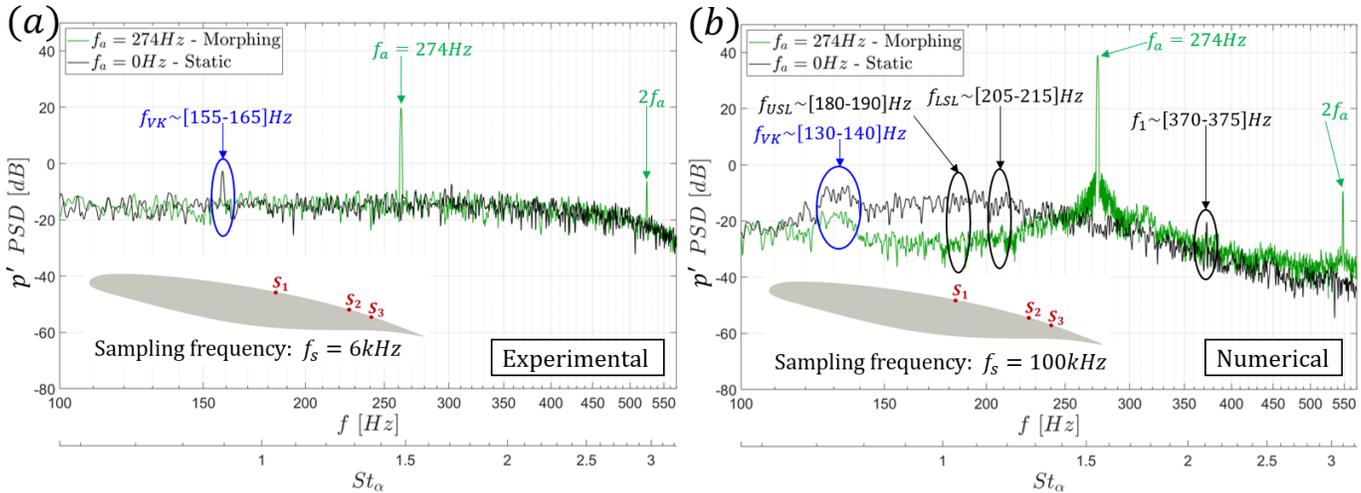

Figure 13 - PSD of pressure fluctuations signals $p'$ on sensor $S_2$ located at $x/c = 0.80$ for the constant actuation frequency at $f_a = 274Hz$ and for the static reference case (3.8s duration signals, $Re = 1 \times 10^6$) - (a) Experiments and (b) Numerical simulations. For interpretation of the references to color in this figure legend, the reader is referred to the web version of this article.

Figures 13a and 13b show the spectral analysis of the pressure signals for the static case and the constant actuation frequency case $f_a = 274Hz$ for experiments and numerical simulations respectively. In both approaches, a sharp peak corresponding to the actuation frequency and its harmonics appear. The Von-Kármán instability, corresponding to a sharp peak in the experiments and a wide bump in the numerical simulations, is illustrated in good agreement between the two approaches. In the experiments, the peak is drastically attenuated in comparison with the static case. For the numerical simulations, the spectral amplitude of the wide bump associated with the Von-Kármán mode is also noticeably attenuated. Furthermore, the instability modes associated with the USL and LSL in the numerical simulations are also modified by morphing, reducing their spectral amplitude in comparison with the static case. This spectral analysis highlights the ability of this type of morphing to attenuate the amplitude of these instabilities and significantly reduce the overall spectral amplitude, (see the frequency range $f = [120\text{-}240]Hz$ in figure 13b where a decay of the order of $10dB$ is obtained). In addition to the aerodynamic performance benefits discussed in detail in the next section, this lowered spectral amplitude is beneficial for reduction of aerodynamic noise sources. This is quite rare in the literature, where devices able to reduce drag, such as active jet vortex generators, generally increase noise.

**4.2.1) Numerical parametric study with respect to the actuation frequency**
*Wake dynamics*

Figure 14 shows streaklines visualisation of the wake for the static reference case and for different constant actuation frequencies. These frequencies were chosen to closely correspond to the different natural frequencies of the flow introduced earlier (section 4.1) in order to highlight the vibration effects in the near and far wake regions. As can be seen in figure 14b, actuating near the Von-Kármán natural frequency enhances the alternating vortex shedding and produces a wider wake in comparison to the static case. Therefore, the local circulation is significantly increased in comparison with static case, but the drag is also drastically increased. At the end, the mean lift-to-drag ratio is decreased together with the significant increase in terms of *rms* on the aerodynamic forces. Actuating close to the Von-Kármán natural frequency is therefore detrimental for the aerodynamic performance.

Actuating close to the USL natural frequency or its first harmonics (figures 14c and 14e) modifies the interactions between the two shear layers, providing an optimal manipulation of the TT/TNT interfaces and taking advantage of the eddy-blocking effect. This optimal actuation delays the appearance of the Von-Kármán mode to a more distant downstream position in the wake and simultaneously shortens the KH formation length $l_{KH}$. As seen in the following section, this actuation produces a decrease in mean drag together with an increase in mean lift, producing a final increase of mean lift-to-drag ratio in comparison to the static case.

Actuating close to the LSL natural frequency has the effect of strongly reinforcing the LSL vortices in comparison with the static case as seen in figure 14d. This drastically increases the local circulation around the airfoil providing the higher increase of mean lift coefficient in comparison to the static case among all the morphing cases investigated. Even if the mean drag coefficient is also increased, the lift increase is more important, providing, finally, an increase in mean lift-to-drag ratio.

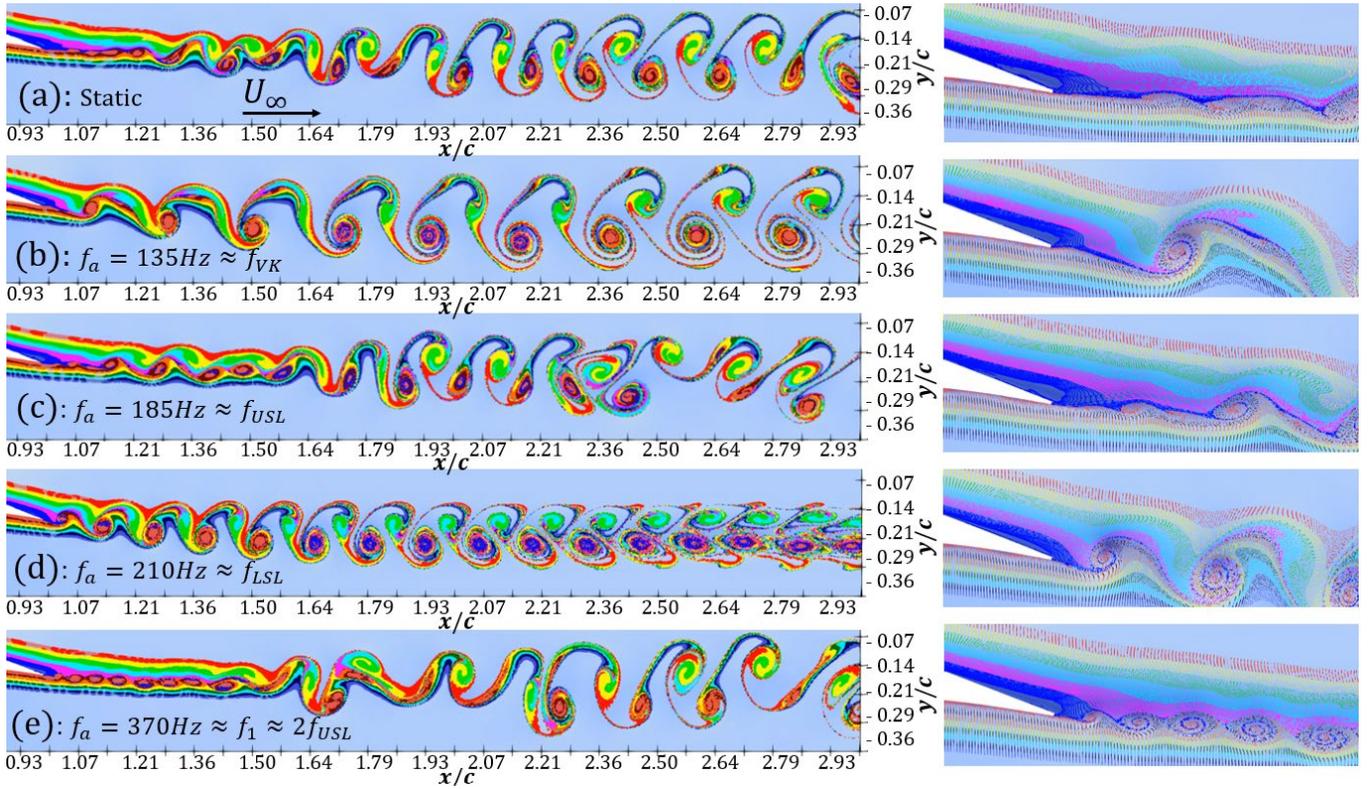

Figure 14 - Streaklines visualisation in the far wake (left) and near trailing edge (right) regions for the static and constant actuation frequency cases - (a): static case, (b): $f_a = 135Hz$, (c): $f_a = 185Hz$, (d): $f_a = 210Hz$ and (e): $f_a = 370Hz$. For interpretation of the references to color in this figure legend, the reader is referred to the web version of this article.

*Aerodynamic performance*

Figure 15 shows the aerodynamic performance in terms of mean and *rms* values for constant actuation frequency in comparison with the static case. Red rectangles highlight interesting regions yielding to an increase of aerodynamic performance. Some of these interesting actuation frequencies are summarized in table 2. An increase of mean aerodynamic performance is targeted, leading to an increase of mean lift-to-drag ratio. This increase can be reached through a decrease of mean drag coefficient and/or an increase of mean lift coefficient. This increase must be accompanied by a *rms* decrease and/or a refraining of *rms* increase. The statistical values (mean and *rms*) were computed consistently for converged statistics during the same physical time of $1.1s$ for all the cases investigated. This duration was chosen long enough to clearly capture the effects of the actuation on the flow and to contain a significant number of coherent structures development.

As shown in table 2 and figure 15, and according to the previous discussion on the wake dynamics (figure 14), actuating at (or close to) the Von-Kármán natural frequency, that is $f_a = f_{VK} \sim 134Hz$, is detrimental for the aerodynamic performance, both in terms of mean lift-to-drag ratio and aerodynamic forces fluctuations. Actuating at (or close to) the USL natural frequency, that is in the range $f_a = f_{USL} \sim [180\text{-}192]Hz$, produces a decrease in the mean drag coefficient together with an increase in the mean lift coefficient, resulting in a final increase in the mean lift-to-drag ratio. However the *rms* is increased in comparison with static case. Actuating at (or close to) the LSL natural frequency, that is in the range $f_a = f_{LSL} \sim [205\text{-}215]Hz$, produces an increase in the mean drag coefficient, but an even greater increase in terms of mean lift coefficient, generating a final increase in mean lift-to-drag ratio in comparison with the static case. However the *rms* is increased in comparison with static case. Actuating at (or close to) the first harmonics of the USL, that is $f_1 \sim 2 f_{USL} = 373 Hz$, is producing a mean drag decrease with a final increase in mean lift-to-drag ratio. Actuating at (or close to) the second harmonic of the USL, $\sim 3 f_{USL} = 568 Hz$, produces the lowest lift *rms* increase in comparison with the static case among all the cases investigated.

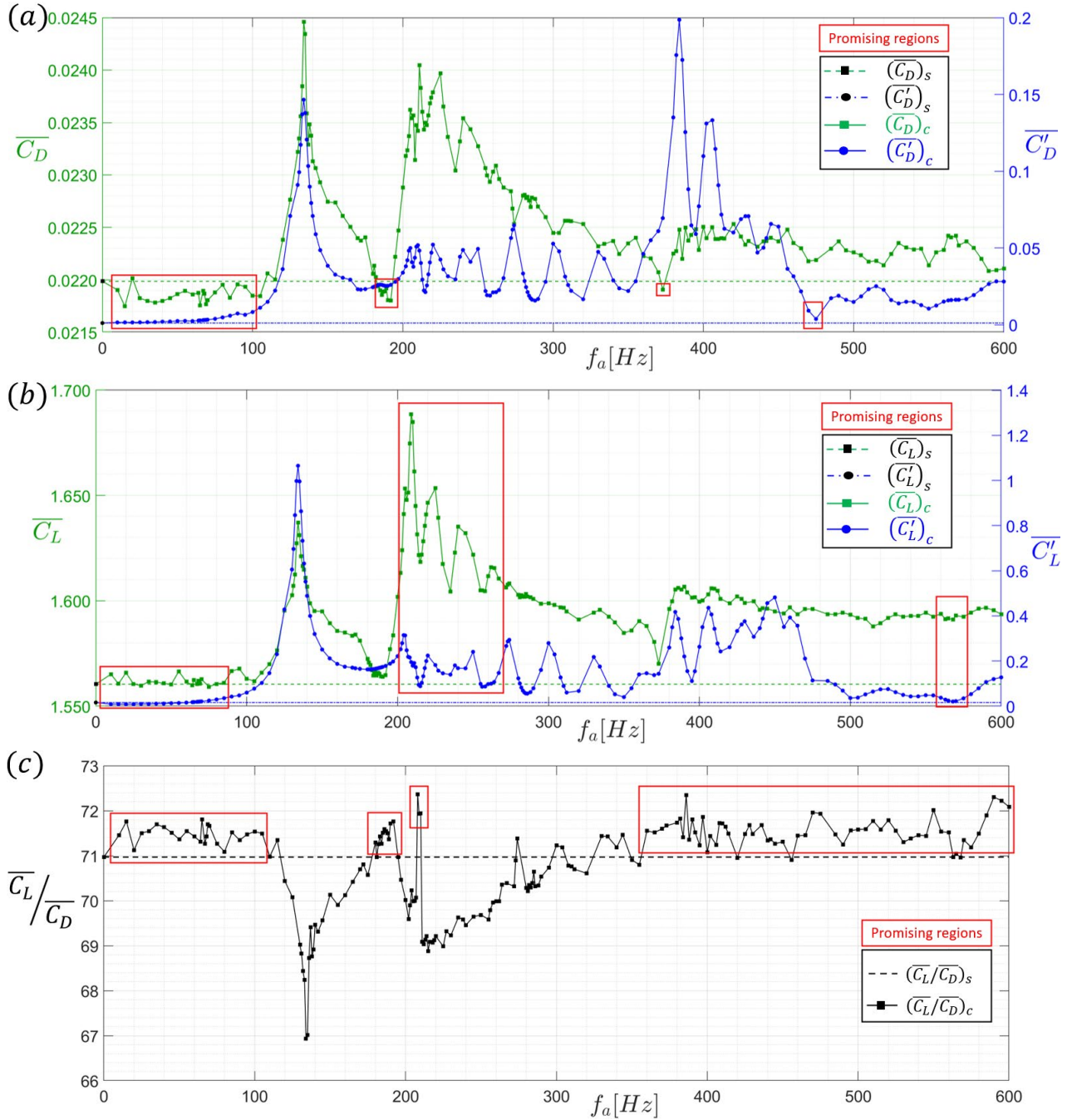

Figure 15 - Aerodynamic performance comparison between constant actuation frequency cases and the reference static case: (a) mean $\overline{C_D}$ and *rms* $\overline{C_D'}$ of the drag coefficient, (b) mean $\overline{C_L}$ and *rms* $\overline{C_L'}$ of the lift coefficient, (c) mean lift-to-drag ratio $\overline{C_L} / \overline{C_D}$. $(X)_s$ and $(X)_c$ subscripts correspond to the mean or *rms* of the forces calculated for the static case and for the constant actuation

frequency respectively. For interpretation of the references to color in this figure legend, the reader is referred to the web version of this article.

| | $f_a$ [Hz] | $\left[\frac{(\overline{C_D})_c - (\overline{C_D})_s}{(\overline{C_D})_s}\right] \times 100\ [\%]$ | $\left[\frac{(\overline{C_L})_c - (\overline{C_L})_s}{(\overline{C_L})_s}\right] \times 100\ [\%]$ | $\left[\frac{(\overline{C_L}/\overline{C_D})_c - (\overline{C_L}/\overline{C_D})_s}{(\overline{C_L}/\overline{C_D})_s}\right] \times 100\ [\%]$ |
|---|---|---|---|---|
| $f_{VK}$ | 134 | +11.3 | +4.92 | -5.69 |
| ~$f_{USL}$ | 185 | -0.39 | +0.37 | +0.76 |
| | 192 | -0.83 | +0.28 | +1.12 |
| ~$f_{LSL}$ | 208 | +5.25 | +7.32 | +1.97 |
| | 210 | +6.52 | +7.98 | +1.37 |
| ~$2f_{VK}$ | 260 | +4.74 | +3.29 | -1.39 |
| | 274 | +2.48 | +3.07 | +0.58 |
| $f_1$~$2f_{USL}$ | 373 | -0.36 | +0.62 | +0.98 |

Table 2 - Summary of the aerodynamic performance: among the large parametric study conducted, only the most interesting cases are presented in this table in relative change (percentage) in comparison with the static case (promising regions included in the red rectangles from figure 15). The most beneficial aerodynamic performance (i.e. mean drag decrease and or mean lift and mean lift-to-drag ratio increase) are underlined in green. For interpretation of the references to color in this table, the reader is referred to the web version of this article.

## 4.3) Wobulation

As previously mentioned, wobulation is a modulation in time corresponding to a continuous sweep of a parameter over a range. Wobulation can be applied to different parameters such as for instance the trailing edge actuation frequency $f_a$ or even the amplitude $a_p$ in a general case. The evolution of each of these parameters can follow a linear or non-linear evolution versus time. In the present study, the amplitude was kept fixed. The scope is to investigate the linear variation of the trailing edge vibration with time. In the following, the time variation of the actuation frequency is referred to as the "*wobulation law*" and is defined as follows:

$$f_w = f_i + \Delta f_w t \quad (4)$$

where $f_w$ is the trailing edge actuation frequency swept by wobulation at a given time $t$, $f_i$ is the initial frequency and $\Delta f_w$ is the frequency increment per second (in $Hz\ s^{-1}$). Figures 16a and 16b show the time evolution of the aerodynamic forces obtained by numerical simulations for an actuation frequency starting from $f_i = 0Hz$ up to $f_w = 445Hz$ with a frequency increment of $\Delta f_w = 50Hz/s$. These parameters are chosen as the reference case for wobulation in the following. A second axis below time shows the linear evolution of the frequency $f_w$. As can be seen in figure 16, wobulation produces a significant modulation of the amplitudes (*rms*) and mean values of the aerodynamic forces. These variations are due to the continuous sweep of the actuation frequency, which produces successive interactions with the different natural frequencies, their sub or superharmonics, generating successive reinforcements through "lock-in" mechanisms or attenuations of the natural frequency effects.

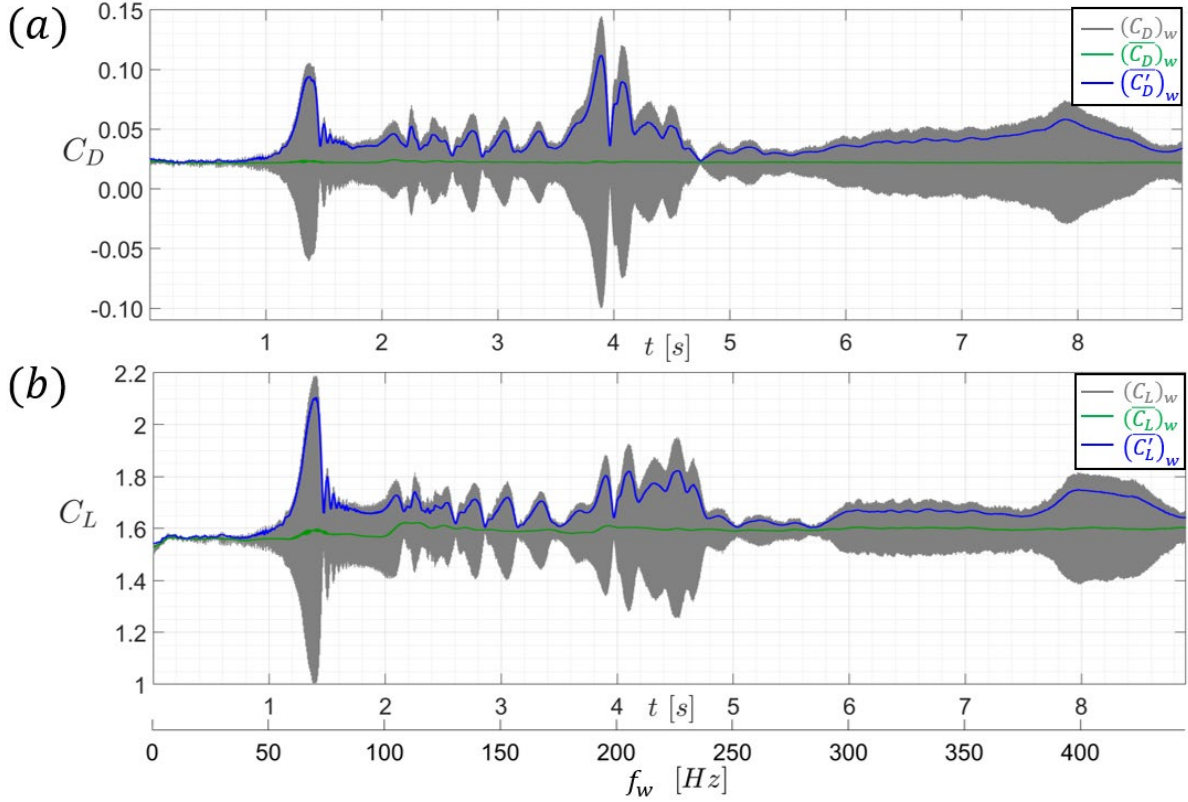

Figure 16 - Aerodynamic forces evolution from numerical simulations (in grey) versus time and time-modulated actuation frequency $f_w$ for one wobulation case ($f_i = 0Hz$ - $\Delta f_w = 50Hz/s$) together with mean (in green) and *rms* (in blue) - (a): drag coefficient $C_D$ and (b): lift coefficient $C_L$. For interpretation of the references to color in this figure legend, the reader is referred to the web version of this article.

### 4.3.1) Time-frequency analysis

A time-frequency analysis was carried out to provide a deeper physical understanding of the aerodynamic forces variation with time as shown above in figure 16. This analysis was conducted on the lift coefficient fluctuation signals $C_L'$, as plotted in figure 17a, using the continuous complex Morlet wavelet transform (Farge 1992). The Morlet wavelet transform is defined as follows:

$$\begin{cases} \psi(t) = \dfrac{1}{2\pi} e^{2i\pi f_0 t} e^{-\left(\frac{t^2}{2}\right)} & (5) \\ C(a,b) = \dfrac{1}{\sqrt{a}} \displaystyle\int_{-\infty}^{+\infty} \sigma(t)\psi^*\left(\dfrac{t-b}{a}\right) dt & (6) \end{cases}$$

where $f_0 = 81.25 \times 10^{-2} Hz$ is the central frequency of the complex Morlet wavelet, $a$ and $b$ are respectively the scaling in frequency and the shift in time parameters of the wavelet, $\sigma(t)$ is the signal, $\psi$ is the wavelet and $C$ is the correlation coefficient. $\psi^*$ and $C^*$ are the complex conjugate of the wavelet and of the correlation coefficient respectively.

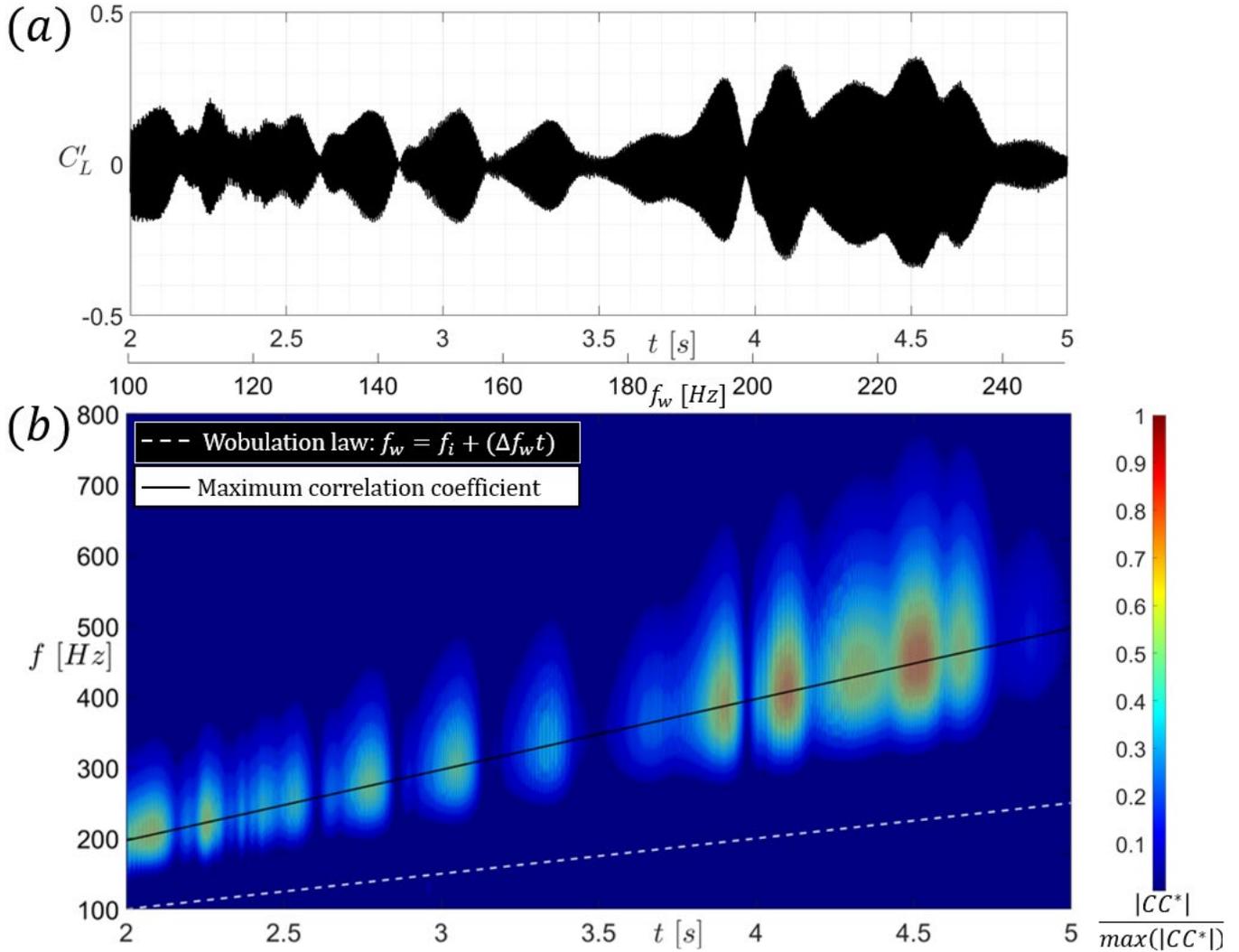

Figure 17 - (a): Lift coefficient fluctuations $C'_L$ signal for the reference wobulation case ($f_i = 0Hz$ and $\Delta f_w = 50Hz/s$) and (b): scalogram of the continuous wavelet transform correlation coefficients $C$. The black line corresponds to the maximum correlation coefficient obtained at each instant $t$ and for each frequency $f$. The absolute value of the wavelet transform coefficients $|CC^*|$ is normalized by the maximum correlation coefficient $max(|CC^*|)$. For interpretation of the references to color in this figure legend, the reader is referred to the web version of this article.

Figure 17b shows the continuous wavelet transform (scalogram) representing the percentage of energy for each correlation coefficient. The absolute value of the wavelet transform coefficients $|CC^*|$ is normalized by its maximum correlation coefficient $max(|CC^*|)$. The white dashed line in the scalogram corresponds to the wobulation law (4) previously introduced, that is the linear evolution of the trailing edge actuation frequency versus time with a slope equals to the frequency increment $\Delta f_w$. The continuous black line corresponds to the wavelet maximum correlation coefficient obtained at each instant $t$ and for each frequency $f$ of the scalogram. This maximum correlation coefficient evolution is found to follow a linear behavior, with a slope equals to twice the increment in all the wobulation cases investigated in this study. It can be introduced as $\alpha_w \Delta f_w$, where $\alpha_w = 2$ is called the *scaling* factor, as explained in the next section 4.3.2. The mathematical derivation of $\alpha_w$ is also given in detail in appendix 1.

Figures 18a and 18b show the spectral analysis of the pressure fluctuation signals for pressure sensor $S_2$ in experiments and numerical simulations respectively. The same initial frequency and increment was chosen between the two approaches, as well as the same range of swept actuation frequencies. The choice of a relatively low increment of $\Delta f_w = 3 Hz/s$ in comparison with the previously introduced increment of $\Delta f_w = 50 Hz/s$ in numerical simulations is due to ensure safety requirements in the actuation of the MFCs during the experiments. A large parametric study of the initial frequency and the increment for numerical simulation is presented in section 4.3.3. The two figures show a good comparison in terms of physical content and order of magnitude of the spectral amplitude. The large bump observed in both approaches and indicated with $f_w = [260\text{-}311]Hz$ corresponds to the actuation frequency range swept by wobulation.

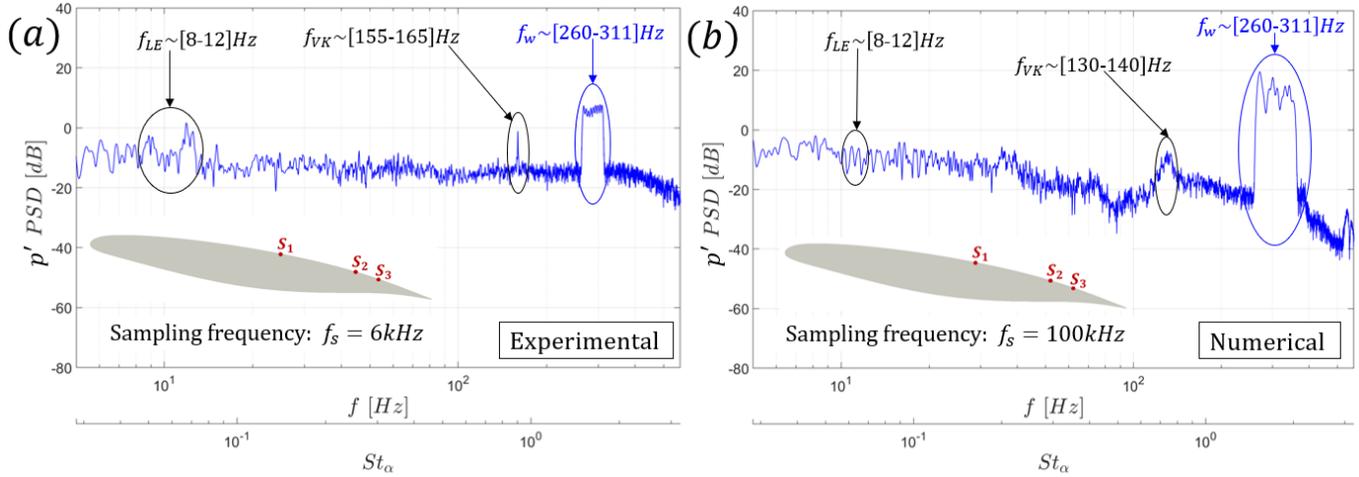

Figure 18 - PSD of pressure fluctuations signals $p'$ on sensor $S_2$ located at $x/c = 0.80$ for wobulation (from $f_i = 260Hz$ up to $f_w = 311Hz$ with a frequency increment of $\Delta f_w = 3Hz/s$, $17s$ duration signals, $Re = 1 \times 10^6$) - (a) Experiments and (b) Numerical simulations.

**4.3.2) Relation between wobulation and constant actuation frequency**

For the sake of clarity, the following analysis focuses on the evolution of the lift coefficient *rms* $\overline{C'_L}$ only, but remains valid for the drag coefficient mean $\overline{C_D}$ and *rms* $\overline{C'_D}$, as well as for the lift coefficient mean $\overline{C_L}$, as will be demonstrated later. As shown in figure 19a, by overlapping the lift coefficient *rms* signal obtained from the wobulation reference case (that is in one sweep over a range of actuation frequencies $f_w$ - figure 16b) and the lift coefficient *rms* signal obtained from a large series of constant actuation frequency cases $f_a$ (previously introduced in figure 15b), the two signals display striking similarities. Indeed, the shape of the blue signal obtained with wobulation appears compressed (or "squeezed") in comparison with the signal obtained with constant actuation frequencies (figure 19a).

Multiplying the range of actuation frequencies swept by wobulation $f_w$ by the scaling factor $\alpha_w$ previously introduced in section 4.3.1, and subtracting a coefficient $\beta_w$ (hereafter referred to as the *shifting factor*) yields to a perfect match between the two signals, as shown in figure 19b. The scaling and shifting factors can be introduced into the proposed "*scaling law*" as follows:

$$f_w^* = \alpha_w (f_w - \beta_w) \quad (7)$$

where $\alpha_w = 2$. The proposed scaling law (7) is the relation between a continuous and linear sweep over a range of actuation frequencies $f_w$ obtained with wobulation and the different constant actuation frequencies $f_a$ previously studied (in section 4.2), such as $f_w^* = f_a$. As explained in the following section 4.3.3, the shifting factor $\beta_w$ was derived by signal correlation analysis after an extensive parametric study conducted on the initial frequency and the increment. It was found to follow a linear dependence on the initial frequency $f_i$ and a non-linear dependence on the frequency increment $\Delta f_w$.

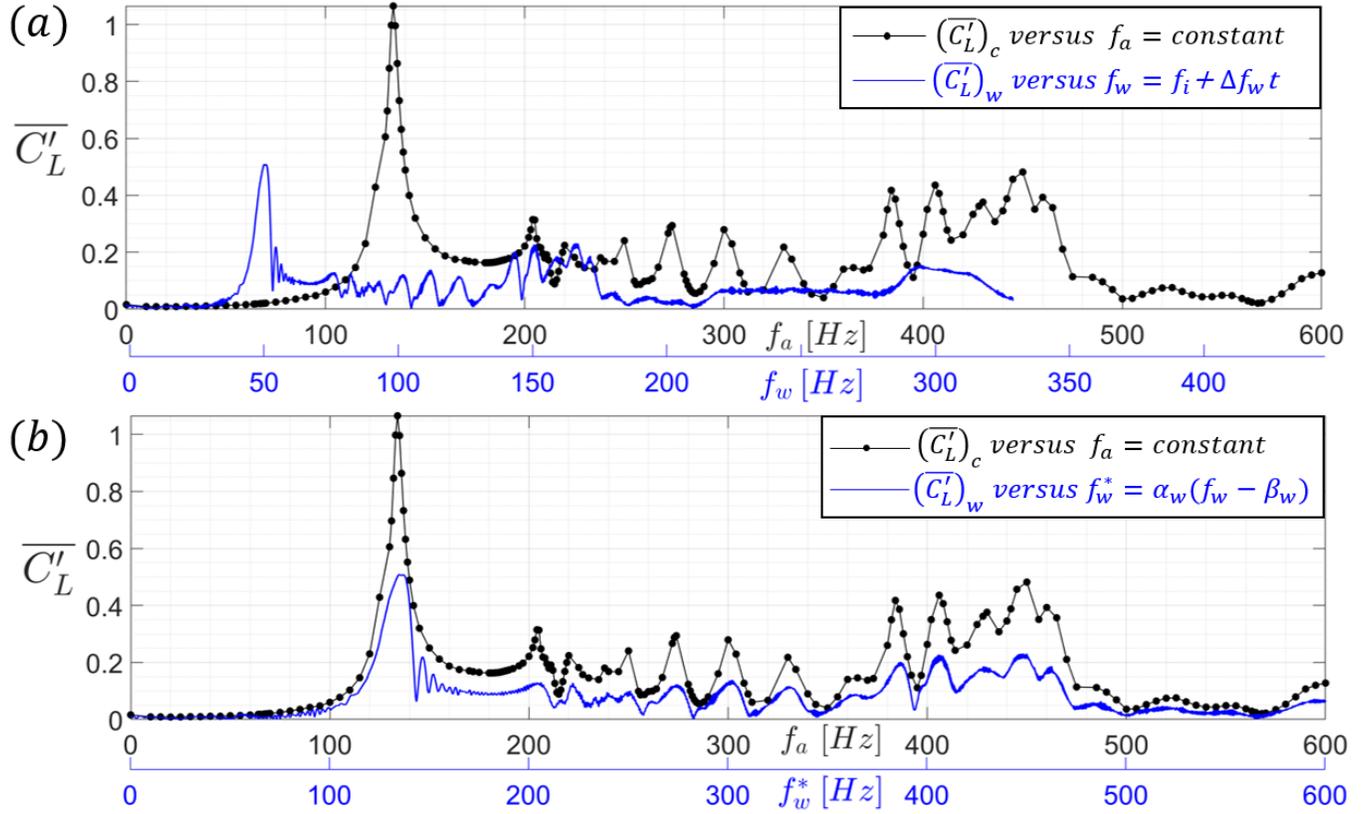

Figure 19 - Lift coefficient *rms* for the constant actuation frequency cases in black and for the reference wobulation case in blue ($f_i = 0Hz$ and $\Delta f_w = 50Hz/s$). Wobulation is plotted versus (a) $f_w$ and (b) $f_w^*$. For interpretation of the references to color in this figure legend, the reader is referred to the web version of this article.

The similarity between the signals obtained with the cases of constant actuation frequency and wobulation is observed not only for the lift coefficient *rms* $\overline{C_L'}$, but also for the drag coefficient mean $\overline{C_D}$ and *rms* $\overline{C_D'}$ and for the lift coefficient mean $\overline{C_L}$, as respectively shown in figures 20a, 20b and 20c. A more quantitative analysis was performed using Pearson's coefficient, which indicates the degree of similarity between two signals, (Benesty, et al. 2009). Correlation coefficients were found to provide a high degree of similarity between the signals obtained with wobulation and those obtained with constant actuation frequency, namely 0.78, 0.97, 0.88 and 0.96 for mean drag $\overline{C_D}$, drag *rms* $\overline{C_D'}$, mean lift $\overline{C_L}$ and lift *rms* $\overline{C_L'}$ respectively.

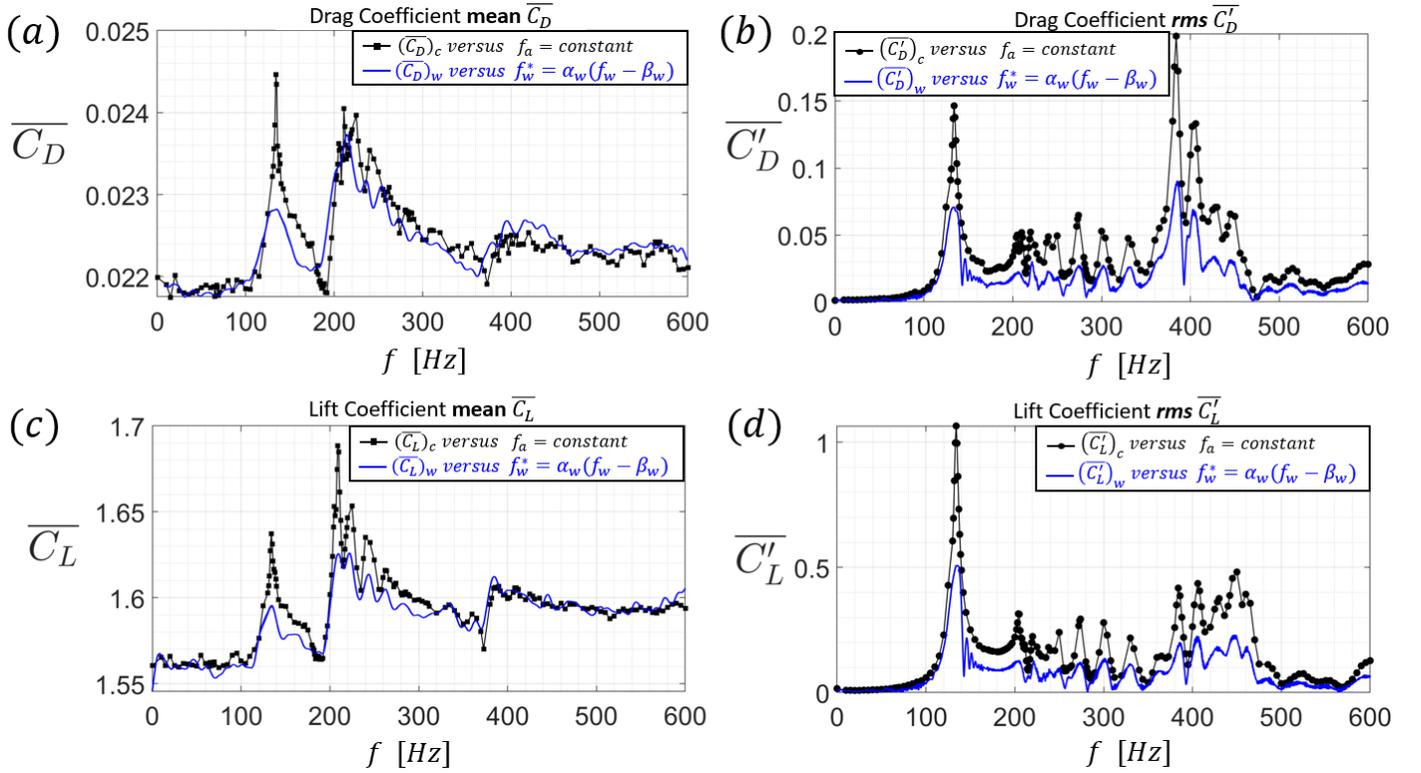

Figure 20 - Mean and *rms* of the aerodynamic forces versus the actuation frequency $f$ for constant actuation cases in black and for one wobulation case in blue ($f_i = 0Hz$ and $\Delta f_w = 50Hz/s$). The same signals are plotted for constant actuation frequency. The wobulation cases are plotted according to the scaling law $f_w^*$ ; (a): mean drag $\overline{C_D}$, (b): drag *rms* $\overline{C_D'}$, (c): mean lift $\overline{C_L}$. For interpretation of the references to color in this figure legend, the reader is referred to the web version of this article.

With a given initial frequency and increment, and knowing the scaling law (7), it is possible to accurately detect optimal constant actuation frequency cases in terms of aerodynamic forces (mean and *rms*). Instead of carrying out a very large parametric study with respect to the constant actuation frequency (see figure 15 in section 4.2 where each point represents a Hi-Fi numerical simulation), only one properly calibrated wobulation allows to efficiently characterize the entire space of parameters. To obtain such a properly calibrated wobulation, it is necessary to understand the effect of the initial frequency and the increment, as discussed in the next section 4.3.3.

### 4.3.3) Influence of the initial frequency and the frequency increment

For clarity reasons and without loss of generality, the analysis focuses only on the evolution of the lift coefficient *rms* $\overline{C_L'}$ signal, but remains valid for the drag coefficient as well. Figure 21 shows the evolution of the lift coefficient *rms* $\overline{C_L'}$ plotted versus the wobulation law (4) for different initial frequencies $f_i$ and increments $\Delta f_w$. The effects of the initial frequency at a constant increment (i.e. $\Delta f_w = 50Hz/s$) and of the increment at a constant initial frequency (i.e. $f_i = 0Hz$) are shown in figures 21a and 21b, respectively.

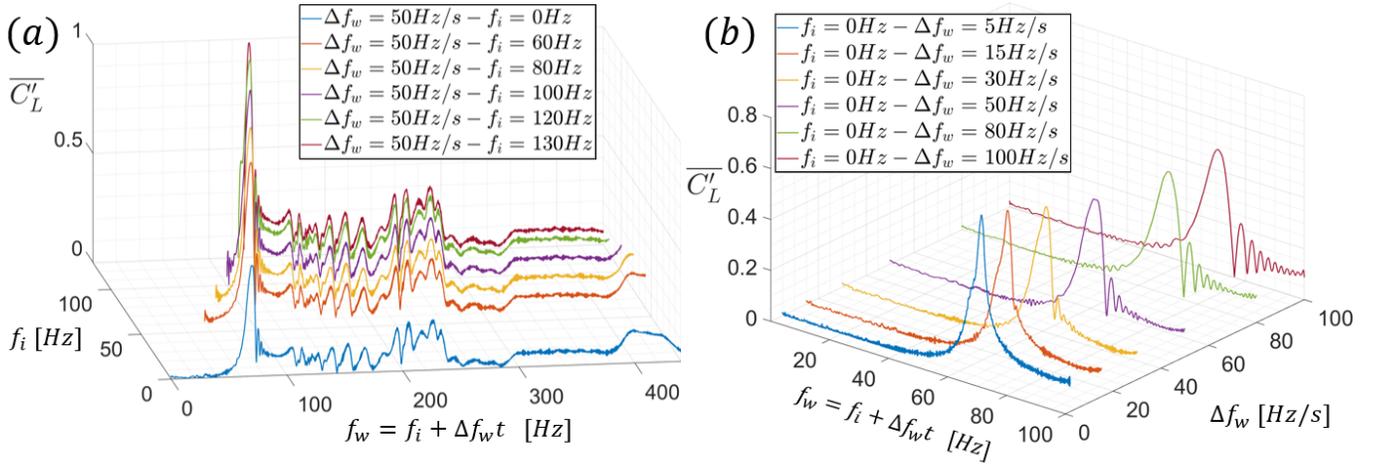

Figure 21 - Lift coefficient *rms* $\overline{C_L'}$ evolution plotted versus wobulation law: (a) for different initial frequencies $f_i$ at a constant increment ($\Delta f_w = 50 Hz/s$) and (b) for different increments $\Delta f_w$ at a constant initial frequency ($f_i = 0Hz$). For interpretation of the references to color in this figure legend, the reader is referred to the web version of this article.

From figure 21a, it is clear that the different initial frequency cases exhibit a strong similarity in the evolution of the lift *rms* $\overline{C_L'}$, but with a shift (or lag) in frequency. Similarly, a lag is observed in figure 21b between the different increment cases plotted. This shift corresponds to the shifting factor $\beta_w$ (in [Hz]), previously introduced in the scaling law (7) in section 4.3.2, and provide a quantitative measure of the lag observed between the different wobulation cases plotted in figure 21. $\beta_w$ is found to depend on both the initial frequency and the increment:

$$\beta_w = f(f_i, \Delta f_w) \quad (8)$$

$\beta_w$ was calculated based on signal correlation between one reference wobulation case (plotted in blue in figures 21a and 21b) and different wobulation signals obtained either with different initial frequencies keeping the increment constant (figure 21a) or with different increments keeping the initial frequency constant (figure 21b). When the initial frequency is varied for the same increment, $\beta_w$ is found to have a linear evolution with respect to the initial frequency. When the increment is varied for the same initial frequency, $\beta_w$ is found to have a non-linear behavior with respect to the increment, following a third polynomial expression as the best fit approximation is given by equation (10):

$$\beta_w = \begin{cases} f(f_i) = A_0 + A_1 f_i & (9) \quad \Delta f_w = cst \\ f(\Delta f_w) = B_0 + B_1 \Delta f_w + B_2 \Delta f_w^2 + B_3 \Delta f_w^3 & (10) \quad f_i = cst \end{cases}$$

where $A_0, A_1, B_0, B_1, B_2$ and $B_3$ are constants, respectively equal to: $A_0 = 2.239$, $A_1 = 0.5$, $B_0 = 0.1196$, $B_1 = -0.5843$, $B_2 = 1.209$ and $B_3 = 2.5134$. Figures 22a and 22b respectively show the linear evolution of the shifting factor $\beta_w$ with respect to the initial frequency for a constant increment of $\Delta f_w = 50Hz/s$ (equation (9)) and the non-linear evolution of $\beta_w$ with respect to the increment for a constant initial frequency of $f_i = 0Hz$ (equation (10)). $A_0$ is the lag due to the increment for which the law is plotted. It is shown that the evolution of $\beta_w$ with respect to the increment follows a quasi-linear law for relatively small increments ($\Delta f_w \leq 10 Hz/s$) and tends towards another linear asymptotic behavior for relatively larger increments ($\Delta f_w \geq 50 Hz/s$).

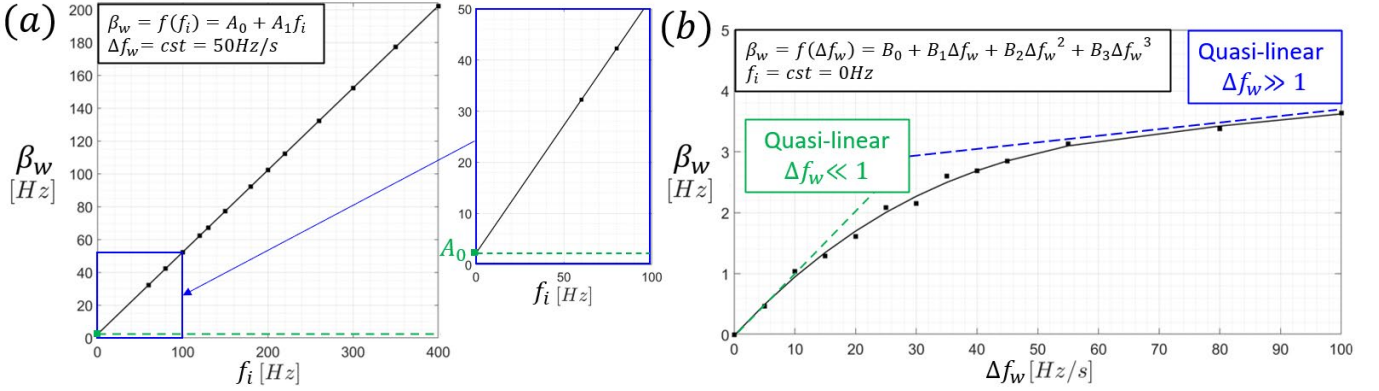

Figure 22 - (a): evolution of the shifting factor $\beta_w$ with respect to the initial frequency $f_i$ with a constant increment $\Delta f_w = 50Hz/s$ and (b) evolution of the shifting factor $\beta_w$ with respect to the increment $\Delta f_w$ with a constant initial frequency $f_i = 0Hz$.

Once the shifting factor $\beta_w$ is known for different initial frequencies and increments, it is possible to plot all the wobulation cases from figure 21 with respect to the scaling law (7) and compare them with the constant actuation frequency cases as shown in figure 23a. Figure 23b shows that the higher the increment, the lower the lift coefficient *rms*.

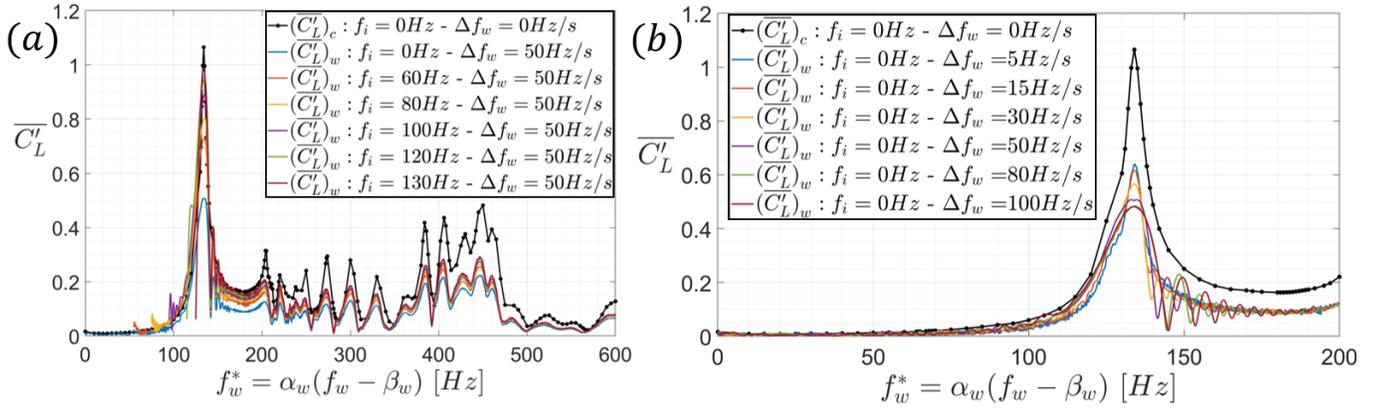

Figure 23 - Lift coefficient *rms* $\overline{(C_L')}_w$ evolution for different wobulation cases plotted with respect to the scaling law $f_w^*$ and overlapped with lift coefficient *rms* $\overline{(C_L')}_c$ for constant actuation frequency cases: (a) for different initial frequencies $f_i$ at a constant increment ($\Delta f_w = 50Hz/s$) and (b) for different increments $\Delta f_w$ at a constant initial frequency ($f_i = 0Hz$).

## 5) Conclusions

This paper investigates the electroactive morphing effects by means of near trailing edge vibration and slight deformation on the Reduced Scale (RS) A320 prototype of the SMS project, at an incidence of 10°, Reynolds number of 1 Million, and for a low-subsonic regime (Mach number of 0.063). First, a thorough physical analysis of the flow dynamics is presented for the static (non-morphing) case. The natural frequencies associated with the main flow instabilities are determined, namely the upper and lower shear layer ones related to Kelvin-Helmholtz (KH) vortices, and the Von-Kármán instability related to vortex shedding. Secondly, the morphing effects at constant actuation frequency and amplitude are examined. The interaction of the actuation frequencies with the natural instability modes is found to play a key role in the modification of the wake dynamics and to produce significant feedback effects on the wall pressure, leading to a strong impact on the aerodynamic coefficients. After an extensive parametric study conducted over a wide range of constant actuation frequencies, it is shown that:

- Actuating at (or close to) the separated shear layers natural frequencies or their harmonics is beneficial for the aerodynamic performance increase.
- Actuating at (or close to) the upper shear layer natural frequency and its harmonics induces a drag reduction ~1% and a slight lift increase, resulting in a final lift-to-drag ratio increase of ~1%.
- Actuating at (or close to) the lower shear layer natural frequency induces a drag increase, but an even more significant lift increase of ~8%, producing a final lift-to-drag ratio increase of ~2%.
- Actuating at (or close to) the Von-Kármán natural frequency results in a significant drag increase, which might be interesting for the landing phase.

In addition to the aerodynamic performance benefits, a reduction of the instability modes is achieved by optimal actuation frequency ranges, as well as a reduction of the spectral amplitude for specific frequency ranges. This results in a more stable dynamic system, robust for the design and with a reduced energy of frequency modes related to aerodynamic noise sources.

Finally, the linear time modulation of the actuation frequency (wobulation), is investigated. By determining the relation between constant frequency vibration and wobulation, it is found that wobulation provides an accurate and efficient methodology to detect optimal constant actuation frequencies (providing an increase in the mean lift-to-drag ratio, a decrease or refrain of increase in the *rms*, and reduction in the amplitude of the instability modes) with only one frequency sweep instead of a large number of different constant vibration cases. It is shown that wobulation is efficient in targeting optimal morphing parameters, thus saving considerable computational time and resources. By providing an accurate open-loop characterization and rapid identification of the dynamical system in real time, wobulation can be used experimentally in ongoing studies to further optimize the closed-loop feedback controller enabling real time optimization.

# Acknowledgements

The authors are grateful to the National Supercomputing Centers CALMIP, CINES, HPC Strasbourg, TGCC-CEA and Rome for the computer resources allocation, as well as the PRACE Supercomputing allocations N°2017174208 and N°2021250049 projects FWING and BEAMW respectively. This work was carried out under the European Commission H2020 N° 723402 research program SMS "Smart Morphing and Sensing for aeronautical configurations" (http://smartwing.org/SMS/EU and https://cordis.europa.eu/project/id/723402) and partly with the contribution of the European Union project TEAMAERO, https://cordis.europa.eu/project/id/860909/fr. This work was also partially sponsored by the Natural Science and Engineering Research Council of Canada (NSERC) Discovery Grant N°166732. The authors are grateful to the "Signal & Image" processing service of IMFT, S. Cazin, M. Marchal, H. Ayroles and F. Bergamme for their contribution to the PIV measuring techniques and data processing, the dSPACE implementation, the aerodynamic balance settings and improvements together with the LAPLACE technical services. The authors warmly thank Dr. C. Döll and P. Mouyon of ONERA-Toulouse for their initial suggestion to wobulate the frequency of the vibrating trailing edge, as well as D. Harribey, Research

Engineer in LAPLACE, for his high contribution to the morphing prototype. Warm thanks are addressed to the IMFT technical services and workshop, C. Korbuly and R. Soeparno for their technical contribution to the wind tunnel experimental setups and to the prototype construction in the SMS project and beyond.

## Appendix 1

*Mathematical derivation of the scaling factor $\alpha_w$*

Using equation (3) (see section 3) modelling the trailing edge displacement of the airfoil, selecting a specific point (such as for instance the trailing edge $x_{TE}$), and replacing the frequency either by a constant actuation frequency $f_a$ or by a time-modulated frequency swept by wobulation $f_w$, yields to:

$$\begin{cases} (y_{TE})_c = h_p sin(2\pi f_a t) & (11) \\ (y_{TE})_w = h_p sin(2\pi f_w t) = h_p sin[2\pi(f_i t + (\Delta f_w t^2))] & (12) \end{cases}$$

Taking the time derivative of the previous two equations (11) and (12) gives:

$$\begin{cases} (\dot{y}_{TE})_c = 2\pi h_p f_a cos(2\pi f_a t) & (13) \\ (\dot{y}_{TE})_w = 2\pi h_p (f_i + 2\Delta f_w t) cos(2\pi f_w t) & (14) \end{cases}$$

After mathematical derivation, the ratio of equations (13) and (14) yields to:

$$\frac{(\dot{y}_{TE})_w}{(\dot{y}_{TE})_c} = \frac{2\pi h_p (f_i + 2\Delta f_w t) cos(2\pi f_w t)}{2\pi h_p f_a cos(2\pi f_a t)} = \left[\frac{f_i + 2\Delta f_w t}{f_a}\right] \frac{cos(2\pi f_w t)}{cos(2\pi f_a t)} = 2 \left[\frac{\frac{f_i}{2} + \Delta f_w t}{f_a}\right] \frac{cos(2\pi f_w t)}{cos(2\pi f_a t)}$$

$$\frac{(\dot{y}_{TE})_w}{(\dot{y}_{TE})_c} = 2 \left[\frac{f_i + \Delta f_w t - \frac{f_i}{2}}{f_a}\right] \frac{cos(2\pi f_w t)}{cos(2\pi f_a t)} = 2 \left[\frac{f_w - \frac{f_i}{2}}{f_a}\right] \frac{cos(2\pi f_w t)}{cos(2\pi f_a t)} \quad (15)$$

Imposing $t = 0s$ in equation (15) and assuming that the shifting factor is mainly due to the lag caused by the initial frequency $\beta_w \sim A_1 f_i = \frac{f_i}{2}$ (see section 4.3.3 where the lag due to the increment is negligible compared to the lag due to the initial frequency), gives:

$$\frac{(\dot{y}_{TE})_w}{(\dot{y}_{TE})_c} = \frac{2\left(f_w - \frac{f_i}{2}\right)}{f_a} = \frac{\alpha_w (f_w - \beta_w)}{f_a} = \frac{f_w^*}{f_a}$$

By identification, $\alpha_w = 2$.

# Glossary

KH - Kelvin-Helmholtz
VK - Von-Kármán
LSL - Lower Shear Layer
USL - Upper Shear Layer
TT - Turbulent-Turbulent interface
TNT - Turbulent-Non turbulent interface
TE - Trailing Edge
LE - Leading Edge
RS - Reduced Scale
tRS - transonic Reduced Scale
PSD - Power Spectral Density
FFT - Fast Fourier Transform
$rms$ - Root Mean Square
AFC - Active Flow Control
HLFC - Hybrid Laminar Flow Control
HFVTE - Higher Frequency Vibrating Trailing Edge
MFC - Macro Fiber Composite
SMA - Shape Memory Alloy
TRPIV - Time-Resolved Particle Image Velocimetry
VSI - Volume Spline Interpolation
TSI - Trans Finite Interpolation
OES - Organised Eddy Simulation
LES - Large Eddy Simulation
ALE - Arbitrary Eulerian Lagrangian
LU-SGS - Lower Upper-Symmetric Gauss-Seidel
NSMB - Navier-Stokes Multi-Block

# Nomenclature

$t$ - time $[s]$
$x$ - chordwise direction $[m]$
$y$ - vertical direction $[m]$
$z$ - spanwise direction $[m]$
$\alpha$ - angle of attack $[°]$
$c$ - chord of airfoil $[m]$
$s$ - wing span $[m]$
$U_\infty$ - upstream airflow velocity $[m/s]$
$P_\infty$ - upstream airflow pressure $[kg/m/s^2]$
$T_\infty$ - upstream airflow temperature $[K]$
$\rho$ - density $[kg/m^3]$
$\mu$ - dynamic viscosity $[kg/m/s]$
$p$ - pressure $[kg/m/s^2]$
$u$ - horizontal velocity $[m/s]$
$v$ - vertical velocity $[m/s]$
$w$ - spanwise velocity $[m/s]$
$E$ - total energy $[kg\,m^2/s^2]$
$q$ - heat flux $[kg/s^3]$
$\tau$ - shear tensor $[kg/m/s^2]$
$Q$ - $Q$ criterion $[s^{-2}]$
$h_p$ - trailing edge actuation semi-amplitude $[m]$
$a_p$ - trailing edge actuation amplitude $[m]$
$L_p$ - patch length $[m]$
$x_p$ - origin of the patch displacement $[m]$
$x_{TE}$ - trailing edge location $[m]$
$y_{TE}$ - trailing edge vertical displacement $[m]$
$\dot{y}_{TE}$ - trailing edge vertical velocity $[m/s]$
$\Delta t$ - time step $[s]$
$f$ - frequency $[Hz]$
$f_s$ - sampling frequency $[Hz]$
$f_a$ - trailing edge constant actuation frequency $[Hz]$
$f_w$ - trailing edge time-modulated actuation frequency $[Hz]$ (wobulation law)
$f_i$ - (wobulation) initial frequency $[Hz]$
$\Delta f_w$ - (wobulation) frequency increment $[Hz/s]$
$f_w^*$ - (wobulation) scaling law $[Hz]$
$\alpha_w$ - scaling factor $[-]$
$\beta_w$ - shifting factor $[Hz]$
$l_{KH}$ - Kelvin-Helmholtz formation length $[m]$
$d_{inst}$ - instability mean vortex size $[m]$
$f_{LSL}$ - Lower Shear Layer natural frequency $[Hz]$
$f_{USL}$ - Upper Shear Layer natural frequency $[Hz]$
$f_{VK}$ - Von-Kármán natural frequency $[Hz]$
$f_{LE}$ - Leading Edge small separation region natural frequency $[Hz]$
$f_1$ - non-linear interaction mode $[Hz]$
$n_1, n_2, n_3$ - n-integer $[-]$
$p_{exp}$ - experimental pressure signal $[kg/m/s^2]$
$p_{num}$ - numerical pressure signal $[kg/m/s^2]$
$C_D$ - drag coefficient $[-]$
$C_L$ - lift coefficient $[-]$
$C_L/C_D$ - lift-to-drag ratio $[-]$
$Ma$ - Mach number $[-]$
$Re$ - Reynolds number $[-]$
$Re_{eff}$ - effective Reynolds number $[-]$
$St$ - nominal Strouhal number (reduced frequency) $[-]$
$St_\alpha$ - effective Strouhal number $[-]$
$St_d$ - dynamic Strouhal number $[-]$
$c\sin(\alpha)$ - airfoil projected area $[m]$
$A_0, A_1, B_0, B_1, B_2, B_3$ - polynomial fit constants $[-]$
$f_0$ - complex Morlet wavelet central frequency $[Hz]$
$a$ - wavelet scaling factor in frequency $[-]$
$b$ - wavelet shifting factor in time $[s]$
$\psi$ - wavelet $[-]$
$C$ - wavelet correlation coefficient $[-]$
$X\,PSD$ - power spectral density $[dB]$
$\dot{X}$ - time derivative
$X^*$ - complex conjugate
$X'$ - fluctuations
$\bar{X}$ - time averaged mean
$\overline{X'}$ - Root Mean Square ($rms$)
$(X)_s$ - value for the static case
$(X)_c$ - value for constant actuation frequency
$(X)_w$ - value for wobulation
$S_1$ - pressure sensor located at $x/c = 0.60$ and $z/c = 0.56$
$S_2$ - pressure sensor located at $x/c = 0.80$ and $z/c = 0.56$
$S_3$ - pressure sensor located at $x/c = 0.85$ and $z/c = 0.56$